\documentclass[12pt]{article}
\usepackage{amsmath}
\usepackage{graphicx}
\usepackage{natbib}
\usepackage{url} 

\newcommand{\blind}{0}

\usepackage[utf8]{inputenc} 
\usepackage[T1]{fontenc}    
\usepackage{hyperref}       
\usepackage{booktabs}       
\usepackage{amsfonts}       
\usepackage{nicefrac}       
\usepackage{microtype}      
\usepackage{amsmath}
\usepackage[top=3cm, right=3cm, left=3cm, bottom=4cm]{geometry}

\usepackage{bbm}
\usepackage{natbib}
\usepackage{subfigure}
\usepackage[justification=centerlast, font=small]{caption}
\usepackage[table]{xcolor}
\usepackage{enumitem}
\usepackage{amssymb}
\usepackage{comment}
\usepackage{amsthm}
\usepackage{titling}
\usepackage{multirow}
\usepackage[flushleft]{threeparttable}
\usepackage{ dsfont }
\usepackage[T1]{fontenc}

\newcommand{\XFILE}{\textsc{xfile} }

\def\T{{ \mathrm{\scriptscriptstyle T} }}
\def\pr{\text{Pr}}

\newtheorem{theorem}{Theorem}

\newtheorem{proposition}{Proposition}

\theoremstyle{definition}

\graphicspath{{art/}}

\newcommand{\frelu}{\text{fReLu}}
\newcommand{\I}{\mathds{1}}

\begin{document}

\def\spacingset#1{\renewcommand{\baselinestretch}%
{#1}\small\normalsize} \spacingset{1}

\def\simind{\stackrel{\mbox{\scriptsize{ind}}}{\sim}}
\def\simiid{\stackrel{\mbox{\scriptsize{iid}}}{\sim}}


\if0\blind
{
  \title{\bf Accelerated structured matrix factorization}
  \author{
  Lorenzo Schiavon\\
    Department of Statistical Sciences, University of Padova,\\
	Via Cesare Battisti 241, 35121 Padova, Italy.\\
    and \\
	Bernardo Nipoti\\
	Department of Economics, Management and Statistics,\\
	University of Milano-Bicocca,\\
	Piazza dell'Ateneo Nuovo 1, 20126, Milano, Italy.\\
 and \\
 Antonio Canale \\
    Department of Statistical Sciences, University of Padova,\\
	Via Cesare Battisti 241, 35121 Padova, Italy.\\
	}
  \maketitle
} \fi

\if1\blind
{
  \bigskip
  \bigskip
  \bigskip
  \begin{center}
    {\LARGE\bf Accelerated structured matrix factorization}
\end{center}
  \medskip
} \fi

\bigskip
\begin{abstract}
Matrix factorization exploits the idea that, in complex high-dimensional data, the actual signal typically lies in lower-dimensional structures. These lower dimensional objects 
provide useful insight, with interpretability favored by 
sparse structures. Sparsity, in addition, is beneficial in terms of regularization and, thus, to avoid over-fitting.
By exploiting 
Bayesian shrinkage priors, we devise a computationally convenient approach for high-dimensional matrix factorization. The dependence between row and column entities is modeled by inducing flexible sparse patterns within factors. The availability of external information is accounted for in such a way that structures are allowed while not imposed.
Inspired by boosting algorithms, 
we pair the the proposed approach with a numerical strategy 
relying on a sequential inclusion and estimation of low-rank contributions, with data-driven stopping rule.
Practical advantages of the proposed approach are demonstrated by means of a simulation study and the analysis of soccer heatmaps obtained from new generation tracking data.

\end{abstract}

\noindent%
{\it Keywords:} dimensionality reduction, heatmap, infinite factor models,  shrinkage priors, side information, soccer tracking data;
\vfill

\newpage
\spacingset{1.5} 
\section{Introduction}
\label{sec:intro}

Embedding, data compression, and low-rank projections in general, are successful approaches that exploit the general idea that, in complex high-dimensional settings, the actual signal typically lies in low-dimensional structures. Herein, we focus on the analysis of high-dimensional matrix data and propose an approach that consists of a novel model and a tailored computational solution for matrix factorization. Matrix factorization techniques are  widespread in statistics and machine learning. Examples include singular value decomposition, principal component analysis, independent component analysis, and, in general, Gaussian linear factor models \citep{roweis1999unifying}. These methods proved to offer a valuable solution to many practical problems arising in a variety of fields, including collaborative filtering \citep{koren2009matrix}, community detection \citep{mao2017mixed},  signal processing \citep{fu2016power}, genomics \citep{stein2018enter}, and ecology \citep{ovaskainen2017make}, among others. 

\begin{figure}[h!t]
	\centering
	\includegraphics[width=.5\textwidth]{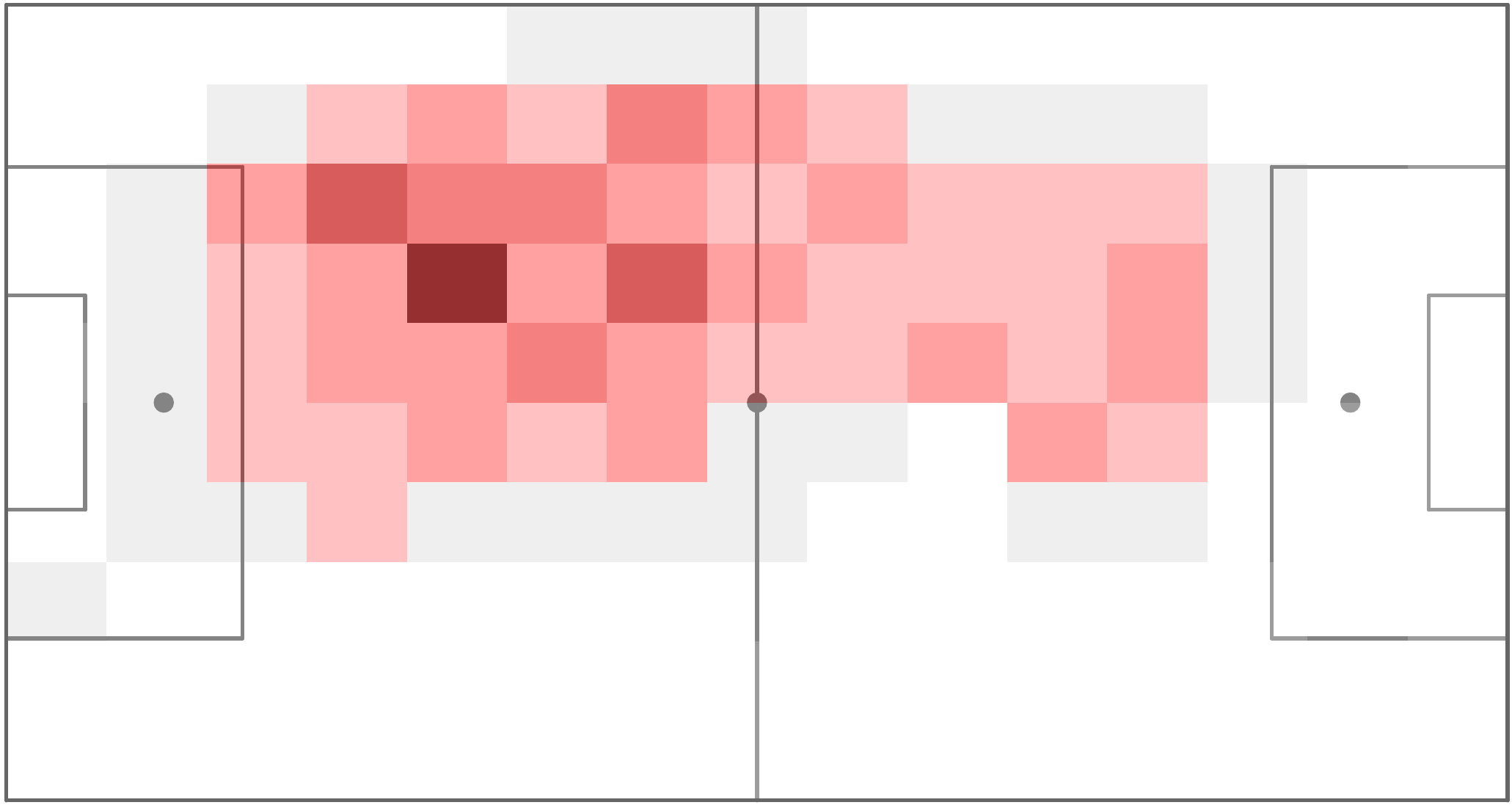}
	\caption{Heatmap for the distance run by a soccer player, attacking from left to right, during the possession time of its team, in different areas of the pitch. Darker areas indicate longer distances, white areas are those not touched by the player.}
	\label{fig:ex-heatmap}
\end{figure}

 We consider a random matrix $Z$ with entries $z_{ij}$, where  $i=1, \dots, n$ and $j=1, \dots, p$ denote subject- and variable-specific indexes, respectively. Data of this sort occur, for example, in recommender systems applications, where user preferences are collected for $n$ users over $p$ items, and in single cell experiments in genomics, 
 with gene expressions recorded for $p$ genes on $n$ cells. Beside these classical examples, the framework of random matrices is well-suited to describe a wide range of data types. In Section \ref{sec:app}, for example, we present the analysis of new generation high-frequency soccer tracking data, by modeling $n$ player-specific heatmaps, each representing the distance covered by a player over the $p$ regions in which the pitch is divided by a given grid. Figure \ref{fig:ex-heatmap} displays one of such heatmaps.

 In matrix factorization models, the random matrix $Z$ is typically factorized as the product of low-rank matrices, e.g. $Z = UV$ where $U$ and $V$ have $k$ rows and columns, respectively, and $k$ is much smaller than $n$ and $p$.  
Low-rank factorizations  have the advantage of reducing the number of unknowns and to provide a parsimonious representation of a possibly high-dimensional matrix. In addition, this dimensionality reduction allows us to link the low-rank factors to interpretable latent constructs, along the lines of the seminal work on factor analysis by \citet{spearman1904general} in psycometrics. 
In our illustrative application, for example,  the vector of distances run by each player 
is represented by a player-specific linear combination of a common set of archetypal heatmaps.

Although, through the years, explainability has often been sacrificed to favor model flexibility and prediction accuracy, by resorting to matrix factorization models that rely on uninterpretable latent factors, 
the recent literature on context-aware matrix factorization models \citep[see, for instance,][]{wu2018collaborative} has put the spotlight back on the importance of providing reasons for the predictions of a recommender system.  
In matrix factorization, this can be accomplished by defining meaningful connections between 
latent factors and context characteristics, user and item traits. In this way, predictions can be interpreted as the consequence of an additive sequence of rules 
accounting for the interactions between user and items, by exploiting the traits that characterize them, and context information. 
In addition, the use of concomitant variables allows one to make predictions about users and items, 
even if in absence of feedback or interactions data, thus avoiding the problem of cold-start recommendations \citep{cortes2020}.
In the recommender systems literature, the inclusion of exogenous information is referred to as context-aware \citep{agarwal2009regression, adomavicius2011context,rendle2011fast,rodriguez2015location} or side information inclusion \citep{porteous2010}. 
Hereafter, we assume that two matrices $X$ and $W$, of dimension $n \times m$ and $p \times q$ respectively, are available, with $X$ storing $q_x$ covariates for each row of $Z$, and $W$ storing $q_w$ metacovariates for each column of $Z$. 
Along the lines of the successful matrix factorization approaches mentioned above, we exploit this auxiliary information to define the latent matrices $U$ and $V$, thus building upon the probabilistic matrix factorization of \citet{mnih2008}.
Specifically, 
we propose a computationally efficient Bayesian hierarchical matrix factorization approach, where the definition of suitable priors allows us to include exogenous information so to favour interpretation and structured regularization of the low-rank latent factors.
Unlike most of the literature on matrix factorization with side information \citep[see, for example,][]{zakeri2018}, we relate the covariates (metacovariates) to the variance of the latent elements rather than their means. We resort to a shrinkage prior, thus favouring  similar patterns of zeros among similar subjects (items), while maintaining a high degree of flexibility. 

The Bayesian perspective allows the uncertainty about the dimension of the latent structures to be dealt with by assigning a prior distribution to the latent rank $k$ \citep{lopes2004}. 
Accordingly, popular Bayesian approaches rely on over-fitted models including more than enough latent elements, with their selection induced by suitable shrinkage priors that adaptively remove unnecessary components by shrinking their coefficients to zero \citep{bhattacharya2011, Legramanti2020, schiavon2022}. 
We propose a simple generalization of the cumulative shrinkage prior of \citet{Legramanti2020} and use it to learn $k$. Allied with this modeling strategy, we devise a 
computational solution for  parameter estimation and prediction, based on posterior maximization. 
This is achieved by means of a computationally efficient strategy based on a forward stage-wise additive procedure, which is the common ground of boosting algorithms \citep{friedman2000, chen2016}. 

The novelty of our proposal is therefore two-fold: it includes a flexible model specification and an efficient algorithm for posterior estimation.
We name the resulting strategy Accelerated Factorization via Infinite Latent Elements, for which we introduce the acronym \XFILE.

Details on the definition and the properties of the hierarchical matrix factorization model we propose are discussed in the next section. Section \ref{sec:comp} describes the steps of the allied computational approach. 
Sections \ref{sec:sim} and \ref{sec:app} illustrate the performance of our proposal when analyzing synthetic and new generation soccer tracking data. Final remarks are presented in Section \ref{sec:discussion}.
Proofs are postponed to the Appendix, while additional figures and details on the algorithm are available as Supplementary Material.


\section{Model and priors}\label{model_prior}
\subsection{Matrix factorization model}
\label{sec:model}

Let $Y$ be a random $n\times p$  data matrix with entries $y_{ij}$, for $i = 1, \dots, n$ and $j=1, \dots, p$.  We assume that each $y_{ij}$ is a transformation of a latent Gaussian variable $z_{ij}$, specifically we set $y_{ij} = f_{ij}(z_{ij})$ for some bijective map $f_{ij}$. For the latent matrix $Z$, with elements $z_{ij}$, we assume the factorization
\begin{equation}
Z = U\Theta V +E,
\label{eq:matrix:fac}
\end{equation}
where $\Theta$ is a $k\times k$ diagonal matrix with diagonal entries $\theta_h$, $U = (u_1, \ldots, u_k)$, $V = (v_1, \ldots, v_k)^{\top}$, where $u_h$ and $v_h$, for $h=1,\ldots,k$, are $n$-variate and $p$-variate column vectors, respectively. The introduction of the diagonal matrix $\Theta$ is discussed in the next section and is equivalent to the core tensor of the so called CP parametrizaton \citep{kolda2009tensor} in tensor factorization models. 
The elements of the $n \times p$ matrix $E$ are independent Gaussian zero-mean errors $\epsilon_{ij}\sim N(0, \sigma_{ij}^2)$. Hence, the associated loglikelihood is  
\begin{equation}
	\log\{\mathcal{L}(Z; U,V,\Theta,\Sigma, X,W)\} = -\sum_{i=1}^{n}\sum_{j=1}^p \big(z_{ij}-\sum_{h=1}^{k} \theta_h u_{ih} v_{hj}\big)^2/\sigma^2_{ij}.
 \label{eq:loglik}
\end{equation}
The opposite of the right-hand side of \eqref{eq:loglik} can be seen as a loss function with weights represented by the parameters $\sigma_{ij}^2$. This is in line with the probabilistic matrix factorization approach of \citet{mnih2008}, where the latent factors $U$ and $V$ appearing in $Z=UV$ are estimated by solving a least squares problem 
obtained by maximizing the posterior distribution of $(U,V)$ in a Bayesian model with Gaussian likelihood
and independent Gaussian priors for $U$ and $V$. We notice that, if a common variance $\sigma^2$ is assumed for the error terms $\epsilon_{ij}$, then the role of $\sigma^2$ is analogous to that of the usual regularization parameter that weights the importance of the penalty function in many machine learning algorithms. Nevertheless, in order to promote a more flexible and robust model specification, 
we allow for varying $\sigma_{ij}^2$, which, in turn, we assume independent from a common prior.

Model \ref{eq:matrix:fac} can be re-written in the form 
\begin{equation}
\label{eq:add-form}
    Z = \sum_{h=1}^k \theta_h u_h v_h^\top + E = \sum_{h=1}^k C_h + E,
\end{equation}
where $C_h$ can be seen as additive rank-one contributions. The additive structure displayed in \eqref{eq:add-form} will be exploited in the algorithm that we will introduce in Section \ref{sec:algorithm}.
In Sections \ref{sec:str-prior} and \ref{sec:rank} we define a hierarchical prior structure for the parameters 
$u_h$, $v_h$ and $\theta_h$, aimed to induce a structured regularization, able to avoid over-fitting and to aid the interpretation of latent structures.

\subsection{Structured shrinkage prior penalty}
\label{sec:str-prior}
As in many contributions in probabilistic matrix factorization, we specify Gaussian priors for the components $u_{ih}$ and $v_{hj}$, of the vectors $u_h$ and $v_{h}$.
Unlike most of the existing literature, however, side information is exploited to model the variances of the Gaussian elements, through shrinkage priors defined as
\begin{equation}
u_{ih} \sim N\{0, \psi_{ih}(x_i)  \}, \qquad v_{hj} \sim N\{0, \phi_{jh}(w_j)\}.
\label{eq:prioretalambda}
\end{equation}
We further assume that $\psi_{ih}(x_i)$ and $\phi_{jh}(w_j)$ are random functions obtained as the product of  idiosyncratic variables
$\tilde{\psi}_{ih} $ and $\tilde{\phi}_{jh}$, discussed later,
and non-linear transformations of linear combinations of covariates and metacovariates. Specifically, we set
$\psi_{ih}(x_i) = \tilde{\psi}_{ih} g_x(x_i^\top \beta_h)^2$ and $\phi_{jh}(w_j) = \tilde{\phi}_{jh} g_w(w_j^\top \gamma_h)^2$, 
where $g_x$ and $g_w$ are non-negative and non-decreasing known functions.
Under these specifications, each $z_{ij}$, for $i=1,\ldots,n$ and $j=1,\ldots,p$, can be represented as 
\begin{equation*}
z_{ij} = \sum_{h=1}^k  g_x(x_i^\top \beta_h) \tilde{\psi}^{1/2}_{ih} \, \tilde{u}_{ih}  \, g_w(w_j^\top\gamma_h) \tilde{\phi}^{1/2}_{jh}\, \tilde{v}_{hj} \, \theta_h+ \epsilon_{ij},
\end{equation*}
where $\tilde{u}_{ih} \sim N(0,1)$ and $\tilde{v}_{hj} \sim N(0,1)$. 
Notably, this construction provides a generalization of Bayesian neural networks \citep{burden2008}. 
A  connection between matrix factorization models and deep learning was already highlighted and exploited by \citet{xue2017}, where the $i$-th row of $U$ and the $j$-th column of $V$
are defined as a low-dimensional mapping of the corresponding row and column of $Y$.
The coefficients $\beta_h$ and $v_h$ in our model, instead, can be thought of as the weights of the two levels of the neural network, with $g_x$ representing the activation function of the hidden layer, typically chosen in the Rectified Linear Unit (ReLu) class of functions. Consistent with these considerations, we consider flexible Rectified Linear Unit (fReLu) functions \citep{qiu2018} as activation functions $g_x$ and $g_w$, i.e. 
$g_x(t) = g_w(t) = \max(t,0)+\varepsilon,$
 with $\varepsilon \geq 0$ fixed, thus guaranteeing a non-decreasing, non-negative, and piece-wise linear behavior, favoring the estimation and the interpretation of the coefficients $\beta_h$ and $\gamma_h$. For the regression coefficients, the following independent priors are assumed:
\begin{align}
	\beta_{1 h} &\sim N(1-\varepsilon, 1), \quad \beta_{d h} \sim N(0, 1), \qquad d=2,\ldots, q, \notag \\
	\gamma_{1 h} &\sim N(1-\varepsilon, 1), \quad
	\gamma_{l h} \sim N(0,1),\qquad l=2,\ldots, m.
 \label{eq:normalprior}
\end{align} 
The parameters $\beta_{1 h}$ and $\gamma_{1h}$ play the role of intercepts that, when no covariates nor metacovariates are available, make the expected contributions associated to $g_x$ and $g_w$, a priori, equal to the multiplicative identity.
Moreover, to catch sparsity patterns within the low-rank contributions, we shrink the noise by setting $\tilde{\psi}_{ih} \sim \text{Ber}(\zeta_n)$ and $\tilde{\phi}_{jh}\sim \text{Ber}(\zeta_p)$, 
with $\zeta_n$ and $\zeta_p$ fixed constants in $(0,1)$. Finally and in line with other contributions in Bayesian factor models \citep[see, e.g,][]{arminger1998}, the prior specification is completed by assuming $\sigma_{ij}^{-2} \simiid \text{Ga}(a_\sigma, b_\sigma)$. We observe that, if the terms $\sigma_{ij}^2$ 
are marginalized out, the errors are marginally distributed according to a central Student-$t$ distribution $t_{2 a_\sigma}(0, b_\sigma/a_\sigma)$, with $2a_\sigma$ degrees of freedom and scale 
$b_\sigma/a_\sigma$.
As a result, the loss function induced by the marginal loglikelihood for $Z$ is the opposite of
\begin{equation}
\label{eq:t-lik}
\log\{\mathcal{L}(Z;U, V, \Theta, a_\sigma, b_\sigma)\}= -\left({a_\sigma+\frac{1}{2}}  \right) \sum_{i=1, j=1}^{n,p} \log \left\{ 1+ \frac{(z_{ij}-\sum_{h=1}^{k} \theta_h u_{ih} v_{hj})^2}{2b_{\sigma}} \right\}.
\end{equation}
Marginalizing with respect to $\sigma^2_{ij}$, 
parameters controlling the level of  penalization in \eqref{eq:loglik}, clarifies the connection between the hyperparameters $a_\sigma$ and $b_\sigma$ and the 
strength of the penalization induced by the prior. Larger values of $b_\sigma$ entail a higher importance of the regularization, while a larger $a_\sigma$ would enhance the role of the loss function in estimating the other parameters. 
%
Given the hierarchical structure of the model, predictions and posterior estimates are less sensitive to small variations of the values of $a_\sigma$ and $b_\sigma$, rather than variations of the elements of $\Sigma$.

\subsection{Rank selection via increasing shrinkage prior}\label{sec:rank}

Rank selection is based on the idea of increasingly shrinking the factor scale $\theta_h$ over the index $h$ so that the rank-one additive contributions $C_h = \theta_h u_h v_h^{\top}$ are negligible for $h$ larger than a certain $k$.
This is accomplished by specifying an over-fitted model with more than large enough rank, combined with an increasing shrinkage prior for the elements of $\Theta$ so to favor a decreasing probability of non-negligible $C_h$ over the index $h$ \citep{bhattacharya2011, Legramanti2020}.
Although a steep decrease of the factor scale over $h$ would certainly induce a low number of relevant factor contributions, the same is likely to lead to a hard-to-interpret model, given the large difference between the magnitude of the 
contributions. It is thus preferable to separate the parameters that control the probability of discarding the negligible contributions and those regulating their magnitude. In addition, it is appealing to define a model that induces a flexible prior distribution on the total number $k$ of active elements. 
All this considered, we adopt and generalize the approach of \citet{Legramanti2020} to define an increasing sequence of truncation probabilities, while maintaining similar scale among contributions. 
Specifically, we consider an ideally infinite sequence $(\theta_h)_{h\geq 1}$ and, for any $h=1,2,\ldots$, we assume $\theta_h = \rho_h \eta_h$, where $\eta_h$ are independent and identically distributed random variables, and $\rho_h$ are independent Bernoulli random variables 
with probability $\pi_h$ of being zero increasing in $h$. The probabilities $\pi_h$ are defined, by means of a stick-breaking construction, as
\begin{equation}
\pi_h= \sum_{l=1}^{h} \varpi_l, \quad \varpi_l= \omega_l\prod_{m=1}^{l-1}(1-\omega_m), \quad \omega_m \simind \text{Be}(1-\delta, \alpha + \delta m),\label{eq:PY}
\end{equation}
with $\delta \in [0,1)$ and $\alpha > -\delta$. The sequence $(\varpi_l)_{l\geq 1}$ follows a two-parameter GEM distribution, distribution characterizing the weights defining the Pitman--Yor process \citep{Pit97}. Definition \eqref{eq:PY} generalizes the one of \citet{Legramanti2020}, which can be recovered by setting $\delta=0$. 
We now focus on the prior distribution induced on the total number $k$ of active elements, starting with its expected value.
\begin{proposition}\label{prop:PY}
For any $h=1,2,\ldots$, we let $\rho_h\simind \text{Ber}(\pi_h)$, with the sequence $(\pi_h)_{h\geq 1}$ defined as in \eqref{eq:PY}, and define $k=\sum_{h=1}^\infty \rho_h$. Then,
\begin{itemize}
\item[i)] if $\delta\in[0,1/2)$,  $\mathbb{E}[k]=(\alpha+\delta)/(1-2\delta)$; 
\item[ii)] if $\delta\in[1/2,1)$,  $k$ has infinite mean.
\end{itemize}
\end{proposition}

Proposition \ref{prop:PY} indicates that, in order to induce shrinking on the number of active elements, one needs to consider $\delta\in[0,1/2)$. In this case, the expected number of $k$ is an increasing function of both $\alpha$ and $\delta$. While one parameter (i.e. setting $\delta=0$) might suffice to model the expected number of active elements, as done in \citet{Legramanti2020}, the availability of a second parameter $\delta$ is convenient if one wants to control, a priori, both expected value and variance of $k$. This can be appreciated by looking at the prior distribution of $k$ for different values of $\delta$ and $\alpha$, as displayed in Figure S1, 
available in the Supplementary Material.

The introduction of the parameters $\rho_h$ thus provides a principled way to select the rank $k$ as the number of non-negligible factors, i.e. every factor $h$ such that $\rho_{h}=1$.
The increasing shrinkage prior elicitation is completed assuming $\eta_h^{-2} \sim \text{Ga}(a_\eta, b_\eta)$. While it is crucial that the priors of the latent contributions are concentrated at zero to effectively shrink small coefficients to zero, it is also important to avoid over-shrinking large signals. The inverse-gamma prior on $\eta_h^2$ implies a power law tail distribution for $\theta_h$, i.e. $\pr(\theta_h>t)\geq at^{-\alpha}$,
for some constants $a > 0$ and $\alpha>0$, and for any $t > M$, with $M$ sufficiently large. This assumption allows us to prove an appealing property of robustness for the rank-one contributions $C_h$,  
as formalized in the next theorem. 

\begin{theorem}
\label{th:robustness}
Let $c_{hij}$ denote a generic entry of $C_h$ and let $\mathcal{P}_{c_{hij}\mid C_{-hij}}(c)$ denote the prior density on $c_{hij}$, conditional on any possible value of $C_1,\ldots,C_{h-1},C_{h+1}, \ldots, C_k$ and the other entries of $C_h$.
Let $\mathcal{L}(Z;C_1, \ldots, C_k)$  denote the likelihood of a factor model specified as in \eqref{eq:add-form}, in terms of the contributions $C_1, \ldots, C_h$.
Assume the following conditions on $\log\{\mathcal{L}(Z;C_1, \ldots, C_k)\}$:
  its first derivative computed with respect to $c_{hij}$ is continuous in $\mathbbm{R}$;
    its second derivative computed with respect to $c_{hij}$ and evaluated at the conditional maximum likelihood estimate $\hat{c}_{hij}$ is, for $\hat{c}_{hij} \rightarrow \infty$, of order greater than or equal to $O(1)$.
If, a priori, $\theta_h$ is power law tail distributed, then
\begin{equation*}
\lim_{\hat{c}_{hij} \rightarrow \infty}    
\left| \hat{c}_{hij}  - \underset{c}{\text{\emph{argmax}}}\left\{ \log \mathcal{L}(Z;c, C_{-hij}) + \log{\mathcal{P}_{c_{hij} \mid C_{-hij}}(c)}\right\} \right| = 0.
 \end{equation*}
\end{theorem}
Under a regular and sufficiently informative loss function, as the one implied by the Gaussian likelihood,  the inverse-gamma prior on $\eta_h^2$ 
guarantees that the maximum of the conditional posterior for $c_{hij}$
is attained close to the estimate $\hat{c}_{hij}$ obtained by minimizing the non-penalized loss function, when $\hat{c}_{hij}$ is large. In other terms, when the data strongly suggest a value far from zero, posterior estimation relies on these only, thus avoiding over-shrinking.

\section{Computational strategy}\label{sec:comp}

\subsection{Forward stage-wise additive maximization}
\label{sec:algorithm}

In the literature on probabilistic matrix factorization, Bayesian point-wise estimates are typically obtained via posterior maximization. The Bayesian specification of the model is exploited to obtain regularized estimates of the parameters by minimizing a loss function penalized by the parameter priors, 
with the probabilistic matrix factorization of \citep{mnih2008} being a notable example. 

We propose a forward stage-wise additive algorithm to approximate the mode of the posterior distribution of the parameters. 
The computational strategy we propose is designed to handle the highly parameterized model we discussed in Section \ref{model_prior}. The algorithm we devise is efficient and constitutes, along with the specification of the prior model, an innovative contribution of our work. The acronym \XFILE, used for Accelerated Factorization via Infinite Latent Elements, refers to the combination of the modeling and the computational strategies we adopt. 

The algorithm we propose relies on the additive representation \eqref{eq:add-form} of the factor model and on the possibility to define the log-prior of the model parameters as the sum of the log-priors of the individual contributions $C_h$, when different factors are assumed independent.
Thus, the model is estimated by sequentially adding a new contribution $C_h = \theta_h u_{h} v_{h}^\top$, where $u_{h}$, $v_{h}$ and $\theta_h$ are the solution of
\begin{equation*}
\underset{(u, v, \theta)}{\text{argmax}} [\log\{{\cal L}(Z; \sum_{l=1}^{h-1} \theta_l u_{l}v_l^\top + \theta u v^\top,  X,W)\}+\sum_{l=1}^{h-1}\log\{\mathcal{P}_{u,v,\theta_l}(u_l, v_l,\theta_l)\}+\log\{\mathcal{P}_{u,v,\theta_h}(u, v, \theta)\}],
\end{equation*}
where $\mathcal{P}_a(\cdot)$ is used to denote the prior probability density function assigned to the random variable $a$.
Conveniently, the optimization problem leading to the definition of the value taken by the $h$-th contribution, is solved with the previous $h-1$ terms held fixed. In other terms, at each iteration of the algorithm, the contribution that most improves the model fit is added.  
Interestingly, a second order approximation of the objective function with respect to $C_{h}$ highlights an insightful connection with gradient boosting procedures \citep{friedman2000}.


This stage-wise approach has two main advantages, especially when the true rank $k$ of the underlying model is small.
First, it allows for fast computations, as the parameter matrices are searched in a space with dimensions that are not larger than necessary. Second, implicit regularization can be performed by stopping the algorithm before convergence, as it is common in boosting algorithms.

The forward stage-wise additive estimation also allows one to provide unique estimates of the otherwise non-identifiable matrices $U$ and $V$.
In fact, given $\sum_{l=1}^{h-1} C_l$, both $u_{h}$ and $v_{h}$ are only identifiable up to an arbitrary rotation $R$ such that $R R^\top=1$. However, such condition is satisfied only by two possible univariate matrices, $R=1$ and $R=-1$. This fact, combined with unimodal and symmetric priors, about zero, for $u_{h}$ and $v_{h}$, leads to only two equally high posterior modes at $(\hat{u}_h, \hat{v}_h)$ and $(-\hat{u}_h, -\hat{v}_h)$, with symmetric interpretation. Convergence of the algorithm is guaranteed when a non-negative constrain on a single element of either $u_{h}$ or $v_{h}$ is fixed. Identification of a unique mode, with contributions naturally ordered according to their fitting capacity, represents an appealing aspect when we are interested in interpreting $U$ and $V$ or the sequential predictive rules induced by $C_h$, as done in the application of Section \ref{sec:app}.

The boosting analogy discussed above sheds new light on the interpretation and use of some parameters. For instance, the parameters $\eta_h$ can be seen as dynamic learning rates for the algorithm, controlling the impact of each step 
\citep{chen2016}. Reducing the impact of a step through a low learning rate makes the 
search for the optimum finer, thus allowing for a better fit. However, as the learning rate gets lower, more steps are needed, making computations slower, and interpretation harder, given that, in our procedure, each step corresponds to an additional factor contribution. The prior on $\eta_h^2$ guarantees sufficient flexibility to balance these two opposite aspects. Morover, setting $b_\eta\leq a_\eta$ ensures sufficient prior mass is assigned to $(0,1)$.

The algorithm stops when it is not possible to increase the log-posterior of the model by adding a factor contribution. This condition can be verified for factor $C_h$ by looking at the value of $\rho_{h}$ that maximizes the log-posterior.
In other terms, while holding $\sum_{l=1}^{h-1} C_l$ fixed, we compare the maximum value $\ell_{\rho_{h}=1}$ of the log-posterior of $\sum_{l=1}^{h} C_l$ under $\rho_h=1$ and $\rho_m=0$, for $m>h$,
and the value $\ell_{\rho_{h}=0}$ obtained by maximizing the log-posterior under $\rho_m=0$, for $m\geq h$, with all the other parameters defining $C_h$ set equal to the corresponding prior modes.
Then, 
we add $C_h$ to the model if $\pr(\rho_h=1)+\ell_{\rho_h=1}>\pr(\rho_h=0)+\ell_{\rho_h=0}$. 
Equations \eqref{eq:pr_delta0} and \eqref{eq:delta_pos} in the Appendix provide closed-form expressions for $\pr(\rho_h=1)$, that is the marginal prior probability that $C_h$ is not shrunk to zero.  

\subsection{Coordinate ascent algorithm for the single contributions}
In order to estimate $(u_h, v_h)$, given the first $h-1$ contributions and that $\rho_h=1$, we rely on a coordinate ascent algorithm \citep{wright2015}. 
At iteration $h$, the loglikelihood for  $Z$ is given by \eqref{eq:t-lik}, with $\tilde{z}_{ij} = z_{ij}- \sum_{l=1}^{h-1} \eta_l u_{il}v_{jl}$ known.
Hence, considering the hierarchical model we defined for the parameters, the goal is to minimize 
\begin{align}
\hspace*{-30pt} &-\sum_{i=1}^n\sum_{j=1}^{p} \left({a_\sigma+\frac{1}{2}}  \right) \log\left\{ 1+ \frac{\left(\tilde{z}_{ij}- g_x(x_i^\top \beta_h) \tilde{\psi}_{ih} \, \tilde{u}_{ih}  \, g_w(w_j^\top\gamma_h) \tilde{\phi}_{jh}\, \tilde{v}_{hj} \, \theta_h \right)^2}{2b_{\sigma}} \right\}\notag\\
&+ \sum_{i=1}^n [\log\{\mathcal{P}_u(\tilde{u}_{ih})\} + \log\{\mathcal{P_\psi}(\tilde{\psi}_{ih})\}]+ \sum_{j=1}^p [\log\{\mathcal{P}_v(\tilde{v}_{hj})\}+\log\{\mathcal{P}_\phi(\tilde{\phi}_{jh})\}]\label{eq:loss_min}\\
&+ \log\{\mathcal{P}_\beta(\beta_h)\}+\log\{\mathcal{P}_\gamma(\gamma_h)\}+ \log\{\mathcal{P}_{\theta_h}(\theta_h)\}.\notag
\end{align}
Notably, the algorithm we propose is not specific to the problem of minimizing \eqref{eq:loss_min} but can be used to deal with any log-posterior, as long as this can be effectively minorized by a quadratic function with respect to the single factor parameters $\tilde{u}_h$ and $\tilde{v}_h$. The computational cost is limited as the algorithm conveniently requires the inversion of only diagonal matrices.

Initial values for the parameters are sampled from their priors. The algorithm then iteratively updates single blocks of parameters, while the others are kept fixed. Within iteration $h$, sub-iterations denoted with the index $t=1,2,\ldots$ are executed until convergence by following the steps summarized below. Additional details are available in the Supplementary Material.

\begin{enumerate}
	\item  \textit{Parameter vector $\tilde{u}_h$ update.}
	Set $\tilde{\psi}_{ih}=1$, for $i=1,\ldots,n$. Then, exploiting the
	minorize-maximize paradigm \citep{wu2010}, update $\tilde{u}_h$ using a quadratic minorant of the Student-\textit{t} loglikelihood, tangent to the current value $\tilde{u}_{h}^{(t-1)}$. To update $\tilde{u}_h$ we rely only on the columns of $\tilde{z}$ such that $\tilde{\phi}_{jh}\neq 0$, since variations on $\tilde{u}_h$ do not change the contribution to the loss function of the columns with index $j$ such that $\tilde{\phi}_{jh}= 0$. 

	\item \textit{Scale vector $\tilde{\psi}_{h}$ update.}
	For $i=1,\ldots,n$, set $\tilde{\psi}_{ih}^{(t)}=1$ if
	the difference between the log-prior of $\tilde{\psi}_{ih}=1$ and the log-prior of $\tilde{\psi}_{ih}=0$ is smaller than the likelihood difference of the two nested models, and 0 otherwise.
	
	\item \textit{Vector $\beta_{h}$ update.}
	The vector $\beta_{h}$ is updated by applying a Newton-Raphson step to maximize the minorant of the Student-\textit{t} loglikelihood tangent to the current value $\beta_{h}^{(t-1)}$. We only consider the observations that carry information on the value of $\beta_h$, that is the observations $ij$ with index $i$ such that $x_i^\top \beta_h^{(t-1)}>0$ and $j$ such that $\tilde{\psi}_{jh} \neq 0$.
	Because of the shape of the $\frelu$ function around zero, the gradient with respect to $\beta_h$ does not exist for some points of the domain. To overcome this issue, we assume the gradient equal to zero if $\frelu(x_i^\top \beta_h)=0$, relying on the subgradient concept \citep{lange2013}.
	
	\item \textit{Vector $\tilde{v}_h$ update.}
	Set $\tilde{\phi}_{jh}=1$ for $i=1,\ldots,n$ and  define $v_h^* = \tilde{v}_h\eta_h$, so that the prior on $v_h^*$, conditionally on $\eta_h$, is the  $p$-variate Gaussian  $N_p(0, \eta_h^2 I_p)$.
	Then, update $v_h^*$ relying on a quadratic minorant of the Student-\textit{t} loglikelihood and on the informative data rows for $v_h$, i.e. any row $i$ such that $\tilde{\psi}_i\neq 0$. Set $\tilde{v}_h^{(t)} = v^*_h/\eta_h$.
	
	\item \textit{Scale vector $\tilde{\phi}_{h}$ update.}
	For $j=1,\ldots,p$, set $\tilde{\phi}_{jh}^{(t)}=1$ if
	the difference between the log-prior of $\tilde{\phi}_{jh}=1$ and the log-prior of $\tilde{\phi}_{jh}=0$ is smaller than the likelihood difference of the two nested models, and 0 otherwise.
	
	\item \textit{Vector $\gamma_{h}$ update.}
    Relying on the subgradient concept, we update $\gamma_h^{(t)}$ following an argument similar to the one adopted for the update $\beta_h$. 
	
	\item \textit{Scale $\eta_{h}$ update.}
	The update of $\eta_h$ exploits the hierarchical specification of $v^*_h$. Given $v_h^* = \tilde{v}_h\eta_h$ with $v_h^* \sim N_p(0, \eta_h^2 I_p)$, the full conditional distribution of $\eta_h^{-2}$ given the other parameters is a Gamma distribution. Thus, the value of $\eta_h^2$ maximizing the objective function is the mode of the inverse-gamma distribution, which is available in closed form.
	
\end{enumerate}

To speed up the algorithm, the number of parameter updates at each iteration could be adaptively reduced.
To accomplish this, \citet{glasmachers2013} propose adaptively changing the frequency of steps occurrence, so to promote the update of the parameters that allow for a larger increase of the objective function. 
We could update more frequently the elements of $\tilde{u}_h$ and $\tilde{v}_h$ corresponding to the elements of $\tilde{\psi}_h$ and $\tilde{\phi}_h$, respectively, equal to $1$ in the latest iteration. 
This approach could be especially beneficial when $n$ and $p$ are very large and when sparsity is expected, i.e. the constants $\zeta_n$ and $\zeta_p$ are set close to zero.

\section{Simulation experiments}
\label{sec:sim}

To illustrate strengths and weaknesses of our methodology, we investigate the predictive ability 
of \textsc{xfile} in different scenarios.  
For comparison, we consider the following methods: the probabilistic matrix factorization \citep{mnih2008}; the collective matrix factorization \citep{cortes2020}; and the content-based model that additively includes exogenous information \citep[see, for instance,][]{porteous2010, zakeri2018} in the low-rank latent matrices. In addition, we also consider, as baseline benchmark, a simple heuristic model that only includes row and column intercepts $b^{(u)}$ and $b^{(v)}$, so that $y_{ij} = b^{(u)}_i + b^{(v)}_j +\epsilon_{ij}$, with $\epsilon_{ij}$ being an error term.
For these methods, we exploit the implementation in the {\sf R} package \texttt{cmfrec} \citep{cmfrec}.

We generate synthetic data from $16$ scenarios defined on the basis of different dimensions, number of factors, types of data generating process, fraction of observed data $\varsigma$, and number of covariates and metacovariates.
For each scenario we simulate $25$ data sets with $n=100$ rows, from $y_{i} = V u_i  + \epsilon_{i}$, with $\epsilon_{i} \sim N_p(0, I_p)$, by letting $(p,k)\in\{(100,7),(500,12)\}$. 
We generate exogenous information matrices $X$ and $W$ as sets of $q_x$ covariates and $q_w$ metacovariates drawn from Bernoulli and Gaussian distributions. We let $q_x \in \{5,10\}$ and set $q_w=q_x$.
The $k$-columns low-rank latent matrices $U$ and $V$ are 
sampled according to the relation with $X$ and $W$, and by following either a multiplicative or an additive structure. In the additive data generating process, the elements of $u_{ih}$ and $v_{hj}$ are drawn from Gaussian distributions with variance equal to $0.25$ and mean defined by a linear combination of $x_i$ and $w_j$, respectively. 
The multiplicative data generating process, instead, follows Equation \eqref{eq:prioretalambda}, with zero-mean $u_{ih}$ and $v_{hj}$ multiplied by a $\frelu$ transformation of linear combinations of covariates and metacovariates. In both processes, sparsity is induced in $U$ and $V$ by randomly setting $75\%$ of the elements equal to zero. 

In every simulated data matrix $Y$, we randomly select a sample of entries $\mathcal{S}$, with $|\mathcal{S}|=\varsigma n p$, letting $\varsigma \in \{0.2,0.4\}$. 
We train each of the competing models by ignoring such entries and, as a measure of their performance, we compute the predictive root mean square error of  model $m$, defined as
$$\text{RMSE}_m=\sqrt{\frac{1}{|\mathcal{S}|}\sum_{l \in \mathcal{S}}(y_l - \hat{y}^{(m)}_l)^2},$$
where $\hat{y}^{(m)}_l$ is the value predicted by  model $m$. 
For \textsc{xfile} we set $\zeta_n=\zeta_p=0.25$,  $\delta=0$, $a_\sigma=b_\sigma$, and $a_\eta=2$. Hyperparameters  $\alpha$, $b_\sigma$ and $b_\eta$ are tuned over a grid of $3\times5\times6$ values and estimated for each scenario based on the predictive RMSE computed in an additional data set with identical simulation settings. 
We use the same data set to tune the most important parameters of the other algorithms. Specifically, for each algorithm, we tune the regularization parameter applied on the squared $L_2$ norms of the matrices over a grid of $11$ values, and the number of latent factors over $\{10,20,30\}$. 

\begin{figure}[t]
    \centering
    \includegraphics[width=.9\textwidth]{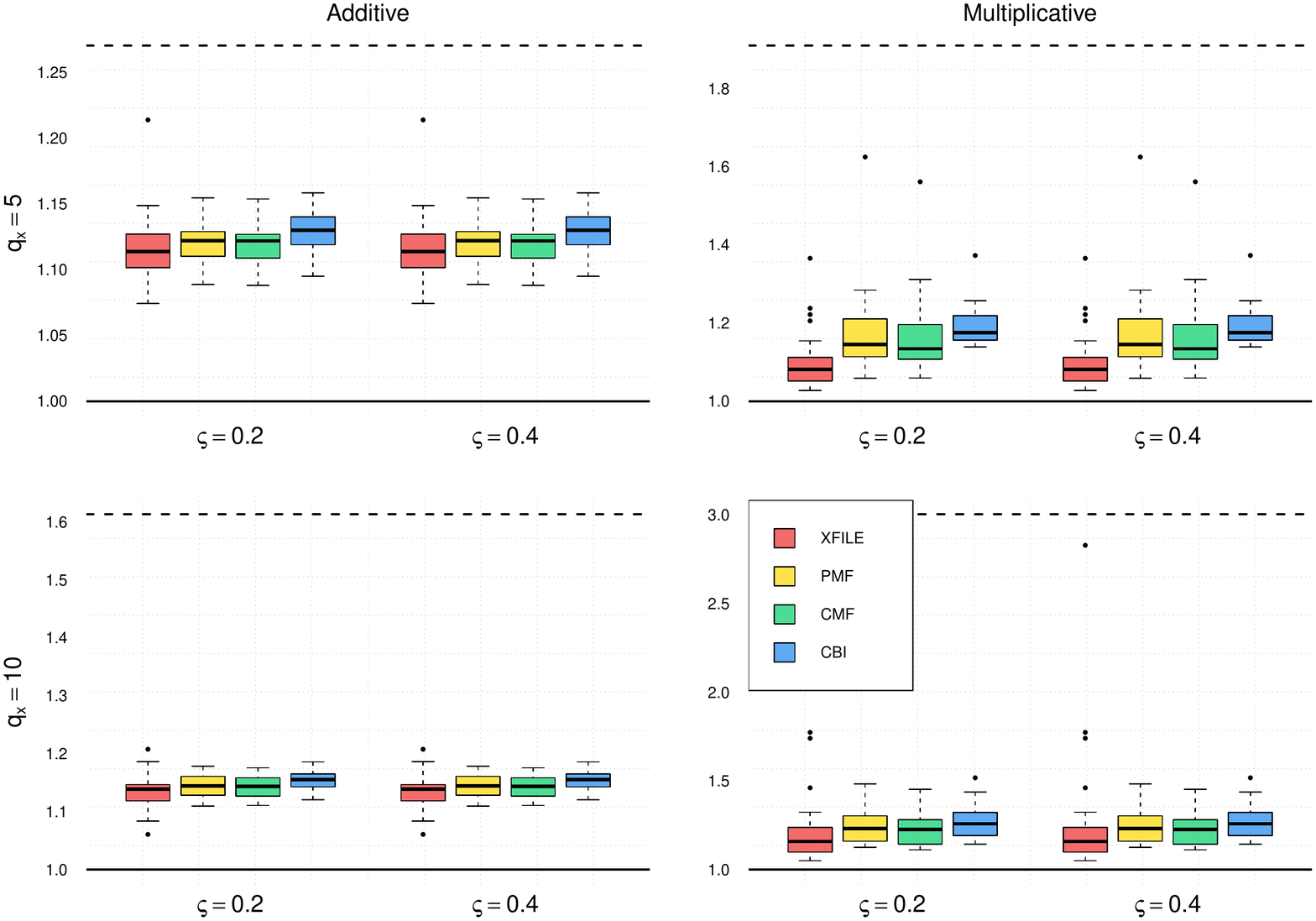}
    \caption{Boxplots of predictive RMSE in 25 replicates under $(n, p,k)=(100, 100, 7)$, varying $q_x$, $\varsigma$, and  data generating process, for \textsc{xfile}, probabilistic matrix factorization (PMF), collective matrix factorization (CMF), and content-based model (CBI). Solid and dashed lines indicate the standard deviation of $\epsilon_{ij}$ and the first quartile of the RMSE of the benchmark.}
    \label{fig:sim7}
\end{figure}

We begin our analysis by considering the scenarios with a moderately small number of items and factors, namely $(p,k)=(100,7)$. Figure \ref{fig:sim7} reports the boxplots of the predictive RMSE for the eight scenarios generated under these settings. The four algorithms perform much better than the baseline benchmark, approach for which,  to enhance the graphical display, we report only a dashed line indicating the value taken by its first quartile. 
The RMSE for \textsc{xfile} is on average smaller than the one of its competitors. Such difference is striking in the multiplicative scenarios, case for which the structure underlying \textsc{xfile} is consistent with the data generating process.
%
\begin{table}[!ht]
\setlength\extrarowheight{-6pt}
    \caption{Median (and interquartile range)  of predictive RMSE in 25 replicates, with $(n,p,k)=(100,500,12)$, different values for $q_x$, $q_w$, $\varsigma$, and two data generating processes. \label{tab:sim12}}
    \vspace{-0.8cm} 
    \centering
    \footnotesize
    \begin{threeparttable}
    \begin{tabular}{rc *{5}{r}}
    \multicolumn{2}{l}{$q_x = q_w = 5$} & & &&&\\[3pt]
    DGP & $\varsigma$ & \XFILE & PMF & CMF & CBI & BB\\
    \hline
    \\[-7pt]
    \multirow[t]{4}{*}{Add.} & \multirow[t]{2}{*}{$0.2$} & $1.134$ & $1.115$ & $1.115$ & $1.112$ & $1.583 $ \\ 
    & & $(1.128 , 1.168 )$ & $(1.110, 1.123 )$ & $(1.109 , 1.123 )$ & $(1.107 , 1.119 )$ & $(1.521 , 1.644 )$ \\ 
    & \multirow[t]{2}{*}{$0.4$} & $1.193$ & $1.153$ & $1.152$ & $1.149$ & $1.598 $ \\ 
    & & $(1.185 , 1.205 )$ & $(1.144 , 1.159 )$ & $(1.146 , 1.158 )$ & $(1.141 , 1.156 )$ & $(1.513 , 1.645 )$ \\ 
    \multirow[t]{4}{*}{Mul.} & \multirow[t]{2}{*}{$0.2$} & $1.135$ & $1.125$ & $1.124$ & $1.120$ & $4.070 $ \\ 
    & & $(1.074 , 1.385 )$ & $(1.109 , 1.140 )$ & $(1.112 , 1.140 )$ & $(1.113 , 1.134 )$ & $(3.183 , 4.675 )$ \\ 
    & \multirow[t]{2}{*}{$0.4$}  & $1.307$ & $1.175$ & $1.175$ & $1.179$ & $4.047 $ \\ 
    & & $(1.177 , 1.645 )$ & $(1.159 , 1.230 )$ & $(1.159 , 1.230 )$ & $(1.161 , 1.207 )$ & $(3.168 , 4.626 )$ \\[10pt]
    \multicolumn{2}{l}{$q_x = q_w = 10$} & &  &&&\\[3pt]
    DPG & $\varsigma$ & \XFILE & PMF & CMF & CBI & BB\\
    \hline
    \\[-7pt]
    \multirow[t]{4}{*}{Add.} & \multirow[t]{2}{*}{$0.2$}  & $1.158$ & $1.156$ & $1.154$ & $1.156$ & $2.205 $ \\ 
    & & $(1.151 , 1.178 )$ & $(1.148 , 1.164 )$ & $(1.145 , 1.161 )$ & $(1.148 , 1.165 )$ & $(2.115 , 2.349 )$ \\ 
    & \multirow[t]{2}{*}{$0.4$} & $1.258$ & $1.215$ & $1.211$ & $1.196$ & $2.227 $ \\ 
    && $(1.246 , 1.272 )$ & $(1.204 , 1.221 )$ & $(1.203 , 1.218 )$ & $(1.187 , 1.206 )$ & $(2.138 , 2.353 )$ \\ 
    \multirow[t]{4}{*}{Mul.} & \multirow[t]{2}{*}{$0.2$} & $1.256$ & $1.175$ & $1.162$ & $1.150$ & $6.769 $ \\ 
    & & $(1.214 , 1.647 )$ & $(1.129 , 1.218 )$ & $(1.138 , 1.215 )$ & $(1.128 , 1.185 )$ & $(5.678 , 8.75 )$ \\ 
    & \multirow[t]{2}{*}{$0.4$} & $1.301$ & $1.292$ & $1.293$ & $1.200$ & $6.614 $ \\ 
    & & $(1.226 , 1.449 )$ & $(1.225 , 1.355 )$ & $(1.225 , 1.355 )$ & $(1.176 , 1.243 )$ & $(5.653 , 8.976 )$ \\
    \end{tabular}
    \begin{tablenotes}
      \footnotesize
      \item DGP: data generating process; PMF: probabilistic matrix factorization; CMF: collective matrix factorization; CBI: content-based model; BB: baseline benchmark.
    \end{tablenotes}
  \end{threeparttable}
\end{table}
Next we consider the scenarios with a larger number of items and factors, namely $(p,k)=(500,12)$. Results are summarized in Table \ref{tab:sim12}, which reports the medians and the interquartile ranges of the predictive RMSE. 
Also in this case, the baseline benchmark performs uniformly worse than all the other algorithms. 
The advantages of \textsc{xfile} appear neutralized, 
with the larger value of the true $k$ leading to a worsening of its performance. Apparently, the ability of \textsc{xfile} to automatically infer the number of latent factors is less and less important as $k$ increases.
In addition, in high signal-to-noise settings, the role of covariates and metacovariates is less apparent, resulting in the possibility of making accurate predictions by finely tuning the parameters of flexibly specified models, while ignoring exogenous information. 

In summary, the simulation study indicates that, when $k$ is moderately small, the stage-wise additive estimation process of \textsc{xfile}, combined with the flexible use of exogenous information, is convenient both in terms of prediction accuracy and interpretation. This is further illustrated with the application presented in the next section. On the other hand, the advantage of using the  \textsc{xfile}, in terms of predictive ability, is mitigated as $k$ gets larger.

\section{Soccer tracking-data heatmaps decomposition}
\label{sec:app}

\subsection{Non-Gaussian distance run heatmaps}
We focus on the analysis of new generation high-frequency soccer tracking 
\if0\blind
{data, provided by 
MathAndSport\footnote{MathAndSport s.r.l. is a sport analytics company based in Italy.} }\fi
\if1\blind{data. }\fi
%
%
Specifically, we consider players' heatmaps, which are graphical representations of the actions of a player over a certain period of time. The pitch is divided into smaller regions by a given grid and the regions are colored depending on the intensity of the action. Figure \ref{fig:ex-heatmap} in the Introduction provides an example of such data.
A heatmap can be described as a $p$-variate vector, where $p$ is the number of areas in which the pitch is divided, and the $j$-th component reports the intensity of the player's action in the $j$-th area. In a soccer match we could collect $n=22$ heatmaps, only by considering the players appearing in the initial line-ups.  
More in general, we are interested in analyzing a collection of heatmaps, which can be represented by a two-dimensional array $Y$. An assumption of independence across either rows or columns of $y$ might be unreasonable, as the intensity of the action of a player in a given region of the pitch is likely to depend on the intensity of the action of the same player in other regions and the one of other players in the same region.  
Observed heatmaps typically display several areas with zero distance covered and positive continuous values elsewhere, which can be encoded into an $n \times p$ heatmap data matrix $Y$ with $y_{ij}\geq0$, for $i=1,\ldots,n$ and $j=1,\ldots,p$.
Recalling the notation of Section \ref{sec:model}, we model the data as a deterministic transformation of an underlying Gaussian model $y_{ij} = f(z_{ij})$, where $Z = U\Theta V + E$, with $E$ being a matrix of independent Gaussian errors $\epsilon_{ij} \sim N(0, \sigma_{ij}^2)$. 
We consider the transformation $y_{ij}=z_{ij} \I_{z_{ij}>0}$, 
which, when the first $h-1$ factors are known, coincides with
\begin{equation*}
	y_{ij}=\begin{cases}
    \sum_{l=1}^{h-1} u_{il} \theta_l v_{jl} +\tilde{z}_{ij} \quad \text{if } \tilde{z}_{ij}>-\sum_{l=1}^{h-1} u_{il} \theta v_{jl} \\
	0 \qquad \qquad  \qquad \qquad \text{if } \tilde{z}_{ij}\leq -\sum_{l=1}^{h-1} u_{il} \theta v_{jl}.
	\end{cases}
\end{equation*}
We treat the Gaussian residual matrix $\tilde{Z}$, with $\tilde{z}_{ij}$ in position $(i,j)$, as an additional factor-specific parameter to update at each additive iteration of the algorithm, 
in order to maximize the posterior. 
We thus add a greedy step in the loop of the algorithm described in Section~\ref{sec:algorithm} to update the matrix $\tilde{Z}$, while holding the other factor-specific parameters fixed.
Conditionally on the first $h-1$ factors, and on $u_{ih}$, $\theta_h$ and $v_{hj}$, the random variable $\tilde{z}_{ij}$ follows a Student-$t$ distribution 
$t_{2 a_\sigma}(u_{ih} \theta_h v_{hj}, b_\sigma/a_\sigma)$. 
The algorithm of Section \ref{sec:algorithm} is thus completed by adding the following step, written by referring to the $t$-th iteration.	
Additional details are available in the Appendix.
\begin{enumerate}
	\setcounter{enumi}{7}
	\item \textit{Gaussian residual $\tilde{Z}$ update.}
	For $i=1,\ldots,n$, $j=1,\ldots,p$, we set $\tilde{z}^{(t)}_{ij}= y_{ij}-\sum_{l=1}^{h-1} c_{lij}$ if $y_{ij}>0$. If $y_{ij}=0$, we set $\tilde{z}^{(t)}_{ij}$ equal to the value that maximizes the conditional distribution of $\tilde{z}_{ij}$ given $y_{ij}=0$ and the values of $U$, $\Theta$ and $V$, which coincides with a truncated Student-$t$ distribution on the interval $(-\infty, -\sum_{l=1}^{h-1} c_{lij})$.
\end{enumerate}

\subsection{Application and results}

We apply  \textsc{xfile} to a dataset $Y$ of $n=106$ heatmaps of different players collected in five matches of a professional soccer league.
Each heatmap is described by a vector of $p=150$ elements corresponding to the $150$ areas in which the pitch is divided. Each element $y_{ij}$ reports the distance covered by the player $i$ within the cell $j$ during the possession time of his team in the match.
Due to data confidentiality agreements, both players and teams have been anonymized. Side information is used to inform the sparsity structure of $U$ and $V$.
To this end, we consider an $n\times q_x$ covariate matrix $X$, defined by considering $q_x=9$ player-specific binary variables describing the expected role of the player, on the basis of the expected line-up made available before the match.
In addition, we exploit a $p\times q_w$ metacovariates matrix $W$, summarizing information on the location of the $p$ areas of the pich, 
with $q_w=4$ location-specific metacovariates, namely the distance from the centre of the pitch, two binary variables indicating which quadrant of the pitch the area belongs to, and an additional binary variable that is equal to one when the area belongs to one of the two boxes and zero elsewhere.

After standardizing the metacovariates matrix, we set $a_\sigma=1$, $b_\sigma=0.3$, $\delta=0$, $\alpha=5$, $a_\eta=4$, 
$b_\eta = 2$, $\zeta_n=0.1$, and $\zeta_p=0.2$.
The algorithm described in Section \ref{sec:algorithm} stops after $4$ iterations, indicating that adding any factor beyond the fourth one does not improve enough the fit.
The structured shrinkage induced by covariates and metacovariates allows us to identify groups of both players and area cells in each factor, by looking at $U$ or $V$, thus leading to an insightful interpretation of the model.
\begin{figure}[h!t]
	\centering
\includegraphics[width=.85\textwidth]{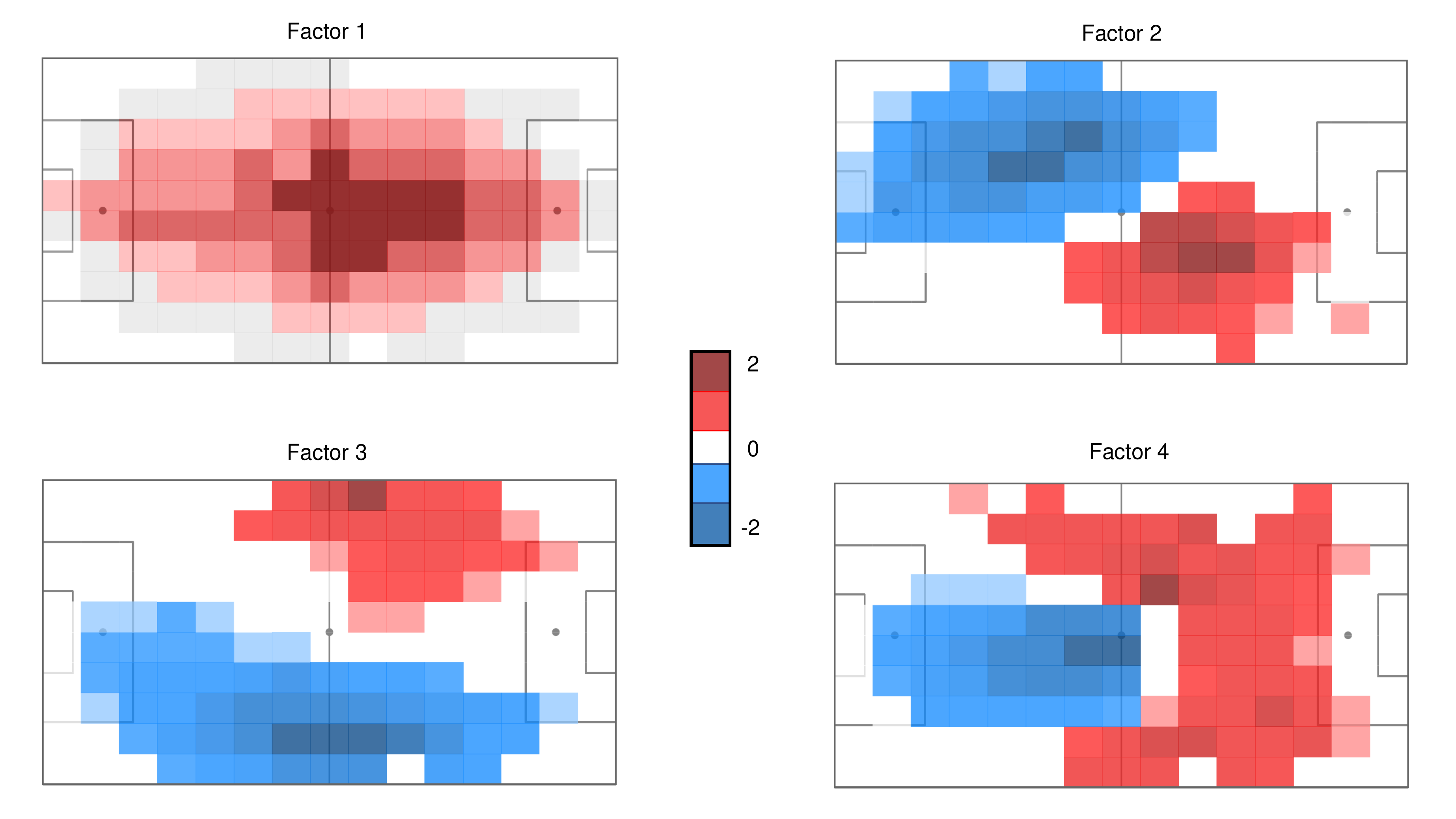} 
	\caption{Heatmaps illustrating the estimate of the four columns of the element-wise product $\tilde{\Phi} \cdot \tilde{V}$. Players attack from left to right.}
	\label{fig:fact-heatmaps}
\end{figure}
Figure \ref{fig:fact-heatmaps} displays the estimate of the four columns of the element-wise product $\tilde{\Phi} \cdot \tilde{V}$, in the form of four heatmaps, where $\tilde{\Phi}$ denotes the $k\times p$ matrix of generic entry $\tilde{\phi}_{hj}$. According to our analysis, a suitable linear combination of such archetypal heatmaps is able to represent sufficiently well any player heatmap of the sample.
The contribution displayed in the top-left corner  of Figure \ref{fig:fact-heatmaps} highlights the areas of the pitch that are mostly involved in the heatmaps, and can be thought of as some sort of baseline heatmap.
The top-right panel shows the second contribution, which helps in distinguishing players who mainly move in the right attacking area, characterized by positive values of $u_2$, from those mostly playing in the left-back, characterized by negative values of $u_2$. If $u_{i2}=0$, the $i$-th player does not follow either pattern.
Analogous considerations apply to the third contribution, shown in the bottom-left corner. The fourth contribution differentiates player behaviors according to a less obvious criterion: 
the blue areas indicate where players involved in the build-up play move, 
while the red areas characterize recurrent movement patterns of players with a stronger attacking propensity, who mostly move on the sides of the pitch when involved in the build-up play.

\begin{figure}[h!t]
	{\centering
	\includegraphics[width=.45\textwidth]{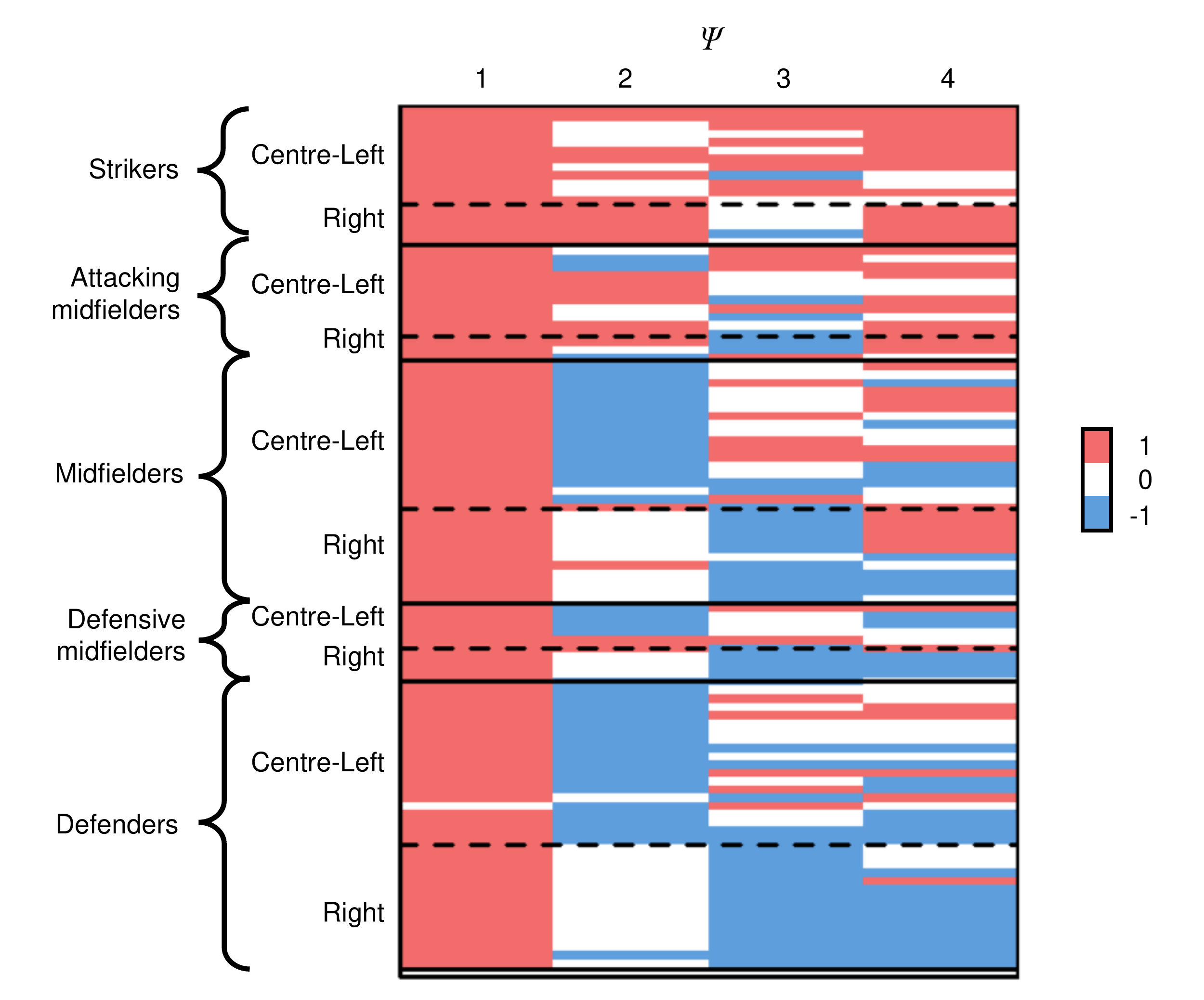} 
\hfill
	\includegraphics[width=.55\textwidth]{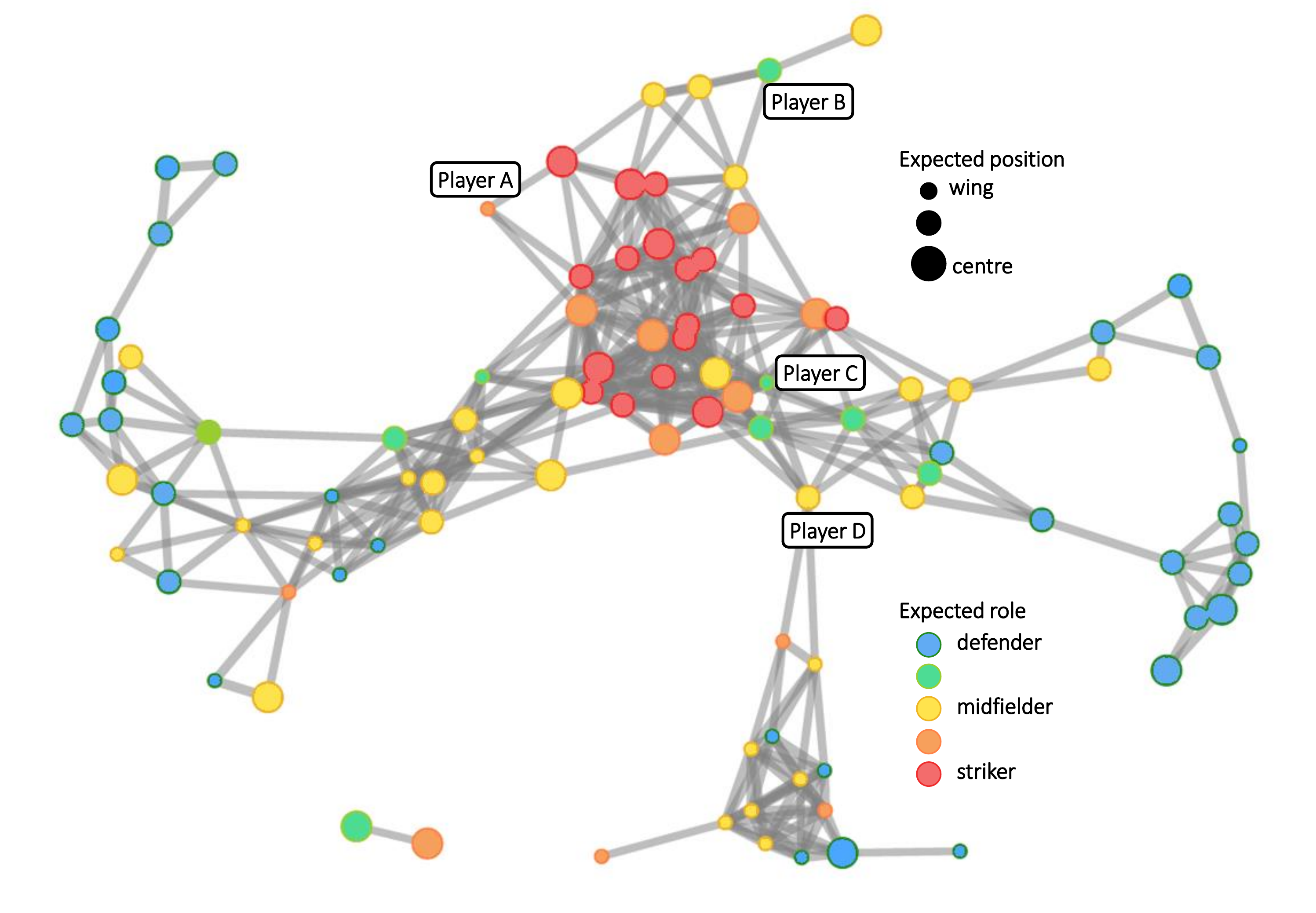} }
	\caption{Left: estimated $\tilde{\Psi} \cdot \text{sign}(U)$, where the rows of the matrix refer to the 106 player heatmaps considered and they are grouped according to their role. Right: network graph representation of the Gaussian kernel similarity of estimated row vectors of $U$.}
	\label{fig:psi}
\end{figure}

Denoting $\tilde{\Psi}$ the $n\times k$ matrix with generic entry $\tilde{\psi}_{ih}$, the estimate of the element-wise product $\tilde{\Psi} \cdot \text{sign}(U)$ is reported in the left panel of Figure \ref{fig:psi}: a colored cell in position $(i,h)$ indicates that the $i$-th player is influenced by the $h$-th contribution. 
As already observed, the first row rank-one contribution acts as a baseline heatmap and turns out to affect almost all the players, with the sole exceptions of the goalkeepers and one defender. 
The second and third columns display blocks characterized by different patterns of colors, 
mostly agreeing with blocks of players with a given role. 
As a by-product, we can easily spot the players who played in a role different from the one they were expected to cover according to the line-up provided before the match. For instance, there are at least three left-side players with null second factor and negative third factor, while the other players on the same side are generally characterized by the opposite behavior. 
Finally, the influence of the last factor appears heterogeneous within each role, especially when defenders and midfielders are considered. This means that our model is also able to identify clusters of players characterized by similar playing styles, regardless of their expected roles. Specifically, the last column of $\tilde{\Psi} \cdot \text{sign}(U)$ identifies three groups according to their propensity in moving forward during the attacking phase.

Each row vector of the estimated matrix $U$ represents the playing style of a single player during a match. Hence, similarities  among different players can be measured by the similarity among the estimated row vectors $u_i$. 
We compute the Gaussian kernel similarity between $u_i$ and $u_l$ as 
$-\exp\{-0.5 (u_i-u_l)^\T \Theta^{-1} (u_i-u_l) \}$.
The right panel of  Figure \ref{fig:psi} shows a network representation based on this similarity metric. It is apparent that players playing in similar roles tend to be clustered together. Interestingly though, defenders appear spread over various clusters, which might be explained by the fact that different teams might have different defending styles. 
While assigned a different expected role, \textit{Player A}, \textit{Player B}, and \textit{Player C} appear close to the cluster of central strikers and attacking midfielders, possibly indicating that these players' playing style was different from what expected before the match. According to football experts, these three professional players are known for their ability to effectively participate in attacking situations while starting from a wider or more defensive nominal position. Finally, it is worth observing that \textit{Player D}, the link connecting different clusters, is universally recognized among soccer insiders for his overall style of play.

\section{Discussion}
\label{sec:discussion}

In line with recent trends on explainable machine learning and context-aware matrix factorization, we introduced a probabilistic matrix factorization approach that includes side information in terms of covariate and metacovariates, leading to interpretable block structures in the latent factors. A key element of our proosal is the computationally efficient algorithm for estimation. 
The latter targets the maximum a posteriori, along the lines of many contributions in the machine learning literature. While MCMC sampling from the posterior distribution is, 
in line of principle, feasible \citep[for example by extending][]{schiavon2022}, the computational cost would be dramatically high, considering the over-parametrized model we defined. A variational Bayes approach based on mean-field approximation \citep[see][]{blei2017} would represent a scalable alternative accounting for full uncertainty quantification. Nevertheless, the characterizing independence assumption between $U$ and $V$ would lead to a degenerate solution for the modes of the low-rank latent matrices, making the interpretation of the same difficult.  

From the practical viewpoint, when compared with methods from the recent literature, the proposed algorithm showed comparable or better performance in scenarios when the true number of latent factors is small or moderate. This is the case, for example, in the soccer tracking data application that motivated our work. The latter application also shows that \textsc{xfile} is able to extract valuable knowledge from high-dimensional data. Given the generality of the proposed approach, we expect these performances to have an impact also in other applied contexts dealing with high-dimensional data matrices. 
We noticed a worsening of the empirical performance of \textsc{xfile} in out-of-sample prediction for high values of the latent rank. A large number of factors, however, entails an intrinsic difficulty in interpretation, which, on the contrary, is one of the motivating characteristics on which we based our proposal.  Notably, a large number of covariates or metacovariates, may lead to possible over-fitting as in standard high-dimensional regression settings. This behaviour could be solved by exploiting suitable shrinkage priors as the Laplace or horseshoe \citep{scott2010} in place of the Gaussians in \eqref{eq:normalprior}. Extensions of the proposed approach to account for these priors are straightforward. 

\if0\blind{
\section*{Acknowledgment}
The authors are grateful to Math\&Sport for providing the data of the application study.
}\fi
\if1\blind{
}\fi

\bigskip
\begin{center}
{\large\bf SUPPLEMENTARY MATERIALS}
\end{center}

\begin{description}

\item[Additional contents:] A detailed description of the steps of the \XFILE algorithm of Section \ref{sec:algorithm}, and additional considerations on its implementation; additional figures of prior distribution on the number of factors (.PDF file) 

\item[R code for \XFILE:] Code to perform the \XFILE algorithm and the simulations of Section \ref{sec:sim}(Zipped .tar file)

\end{description}

\bibliographystyle{jasa3}
\bibliography{biblio}

\newpage
\appendix

\section*{Appendix}


\subsection*{Proof of Proposition \ref{prop:PY}}

\begin{proof}
We start by observing that
\begin{equation}\label{eq:series}
    \mathbb{E}[k]=\mathbb{E}\left[\sum_{h=1}^\infty \rho_h\right]=\sum_{h=1}^\infty \mathbb{E}[\rho_h]=\sum_{h=1}^\infty \mathbb{E}[\mathbb{E}[\rho_h\mid \pi_h]]=\sum_{h=1}^\infty \mathbb{E}[1-\pi_h]
\end{equation}
where, for the second equality, we used Fubini-Tonelli theorem. Moreover,
\begin{equation*}
    \mathbb{E}[1-\pi_h]=1-\sum_{l=1}^{h}\mathbb{E}\left[\omega_l\prod_{m=1}^{l-1}(1-\omega_m)\right]=1-\sum_{l=1}^{h}\mathbb{E}\left[\omega_l\right]\prod_{m=1}^{l-1}\mathbb{E}\left[1-\omega_m\right].
\end{equation*}
We observe that 
$\mathbb{E}[\omega_l]=(1-\delta)\{1+\alpha+\delta(l-1)\}^{-1}$, $\mathbb{E}[1-\omega_m]=(\alpha+\delta m)\{1+\alpha+\delta(m-1)\}^{-1}$,
and write
\begin{align*}
    \mathbb{E}\left[1-\pi_h\right]=1-\sum_{l=1}^h \frac{1-\delta}{1+\alpha+\delta(l-1)} \prod_{m=1}^{l-1} \frac{\alpha+\delta m}{1+\alpha+\delta(m-1)}.
\end{align*}

\begin{itemize} 
\item[i)] The case $\delta=0$ is studied in \citet{Legramanti2020}, and simplifies to 
\begin{equation}\label{eq:pr_delta0}
\pr(\rho_h=1)=\mathbb{E}[1-\pi_h]=\alpha^h/(1+\alpha)^h.
\end{equation}
Thus,
\begin{equation}\label{eq:exp_kapp_case0}
    \mathbb{E}[k]=\sum_{h=1}^\infty\frac{\alpha^h}{(1+\alpha)^h}=\alpha.
\end{equation}
When $\delta\in(0,1)$, with some algebra we obtain
\begin{equation}
    \pr(\rho_h=1)=\mathbb{E}[1-\pi_h]=\frac{\Gamma\left(h+1+\alpha/\delta\right)\Gamma\left((1+\alpha)/\delta\right)}{\Gamma\left(h+(1+\alpha)/\delta\right)\Gamma\left(1+\alpha/\delta\right)}\label{eq:delta_pos}
\end{equation}
which, if $\delta\in(0,1/2)$, leads to
\begin{equation}\label{eq:exp_kapp_case1}
    \mathbb{E}[k]=\frac{\Gamma\left(1/\delta-2\right)\Gamma\left(\alpha/\delta+2\right)}{\Gamma\left(1/\delta-1\right)\Gamma\left(\alpha/\delta+1\right)}=\frac{\alpha+\delta}{1-2\delta}.
\end{equation}
By combining \eqref{eq:exp_kapp_case0} and \eqref{eq:exp_kapp_case1} we conclude that, if $\delta \in[0,1/2)$, then $\mathbb{E}[k]=(\alpha+\delta)/(1-2\delta)$.
\item[iii)] If $\delta\in[1/2,1)$, then \eqref{eq:delta_pos} holds and the series in \eqref{eq:series}
does not converge, thus $k$ has infinite mean.
\end{itemize}
\end{proof}

\subsection*{Proof of Theorem \ref{th:robustness}}

\begin{proof}

We firstly need to prove that the prior distribution of $(\theta_h \mid u_{ih}, v_{hj}, C_{-hij})$ is power law distributed. 
It can be shown by demonstrating that the conditional prior on $( C_{-hij} \mid \theta_h, u_{ih}, v_{hj})$ goes to zero slower than $d\theta^{-\alpha}$ for certain $d,\alpha>0$ and $\theta \rightarrow \infty$, and then applying the Bayes theorem and relying on the power law tail condition on the marginal prior of $\theta_h$.
Let $\mathcal{P}_{x\mid t}(x)$ indicate the prior on $x$ conditionally on $t$. Then, 
\begin{align*}
   \mathcal{P}_{C_{-hij} \mid \theta_h, u_{ih},v_{hj}}(C;\theta_h)&= \mathcal{P}_{C_{h} \mid \theta_h, u_{ih},v_{hj}}(C;\theta_h)
    \prod_{l\neq h} \mathcal{P}_{C_{l}}(C)\\
    &\propto \prod_{l\neq i, m \neq j} \mathcal{P}_{c_{hlm}\mid\theta_h}(c;\theta_h) \prod_{m \neq j} \mathcal{P}_{c_{him}\mid \theta_h, u_{ih}}(c;\theta_h)  \prod_{l \neq i} \mathcal{P}_{c_{hlj}\mid \theta_h, v_{hj}}(c;\theta_h)
\end{align*}
If $c=u_{ih} v_{hj} \theta_h$,  $\mathcal{P}_{c_{hij}\mid \theta_h, u_{ih},v_{hj}}(c)=1$. 
We can write $\mathcal{P}_{c_{hlj}\mid \theta_h, v_{hj}}(c) = \mathcal{P}_{u_{ih}}\{c (v_{hj}\theta_h)^{-1}\} (v_{hj} \theta_h)^{-1}$ and $\mathcal{P}_{c_{hlj}\mid \theta_h, u_{ih}}(c) = \mathcal{P}_{v_{hj}}\{c (u_{ih}\theta_h)^{-1}\} (u_{jh} \theta_h)^{-1}$.
By construction, $\mathcal{P}_{u_{ih}}(u)$ and $\mathcal{P}_{v_{hj}}(v)$ are symmetric around zero and strictly positive when $u$ and $v$ are equal to zero, respectively. Then, $\mathcal{P}_{c_{hlj}\mid \theta_h, u_{ih}}(c;\theta_h) \mathcal{P}_{c_{hlj}\mid \theta_h, v_{hj}}(c) \geq d \theta_h^{-2}$ for a certain $d>0$ when $\theta_h \rightarrow \infty$, ensuring that $(\theta_h \mid u_{ih}, v_{hj}, C_{-hij})$ is power law tail distributed.

If $(\theta_h \mid u_{ih}, v_{hj}, C_{-hij})$ is power law tail distributed, then $(c_{hij} \mid C_{-hij})$ is power law tail distributed by applying the result in Lemma 3 reported in the Appendix of \citet{schiavon2022}.
In other terms, we can write $\mathcal{P}_{c_{hij} \mid C_{-hij}}(c)\geq d |c|^{-\alpha}$ for certain $d, \alpha$ positive constants and $|c| > L$ sufficiently large, or, equivalently, $\mathcal{P}_{c_{hij} \mid C_{-hij}}(c) =  d |c|^{-\alpha}\{1+f(|c|)\}$, with $f(|c|)$ a positive function such that $f(|c|) < |c|^{\alpha}$ when $|c|$ goes to $\infty$.
Hence,
\begin{equation*}
    \frac{\partial \, \log \mathcal{P}_{c_{hij} \mid C_{-hij}}(c)}{\partial c} =  -\alpha |c|^{-1} + \{1+f(|c|)\}^{-1} \frac{\partial \, f(|c|) }{\partial c},
\end{equation*}
with derivative of $f(|c|)$ positive or going to $0$ when $|c| \rightarrow \infty$.
Since $\mathcal{P}_{c_{hij} \mid C_{-hij}}(c)$ is decreasing when $|c| \rightarrow \infty$, the derivative of the logarithm is non positive, allowing one to conclude that
\begin{equation}
\label{eq:lim-score}
    \lim_{c \rightarrow \infty} \frac{\partial \, \log \mathcal{P}_{c_{hij} \mid C_{-hij}}(c)}{\partial c} = 0.
\end{equation}

The mode of the conditional posterior density of $c_{hij}$ is $\tilde{c}_{hij}$ such that
\begin{equation*}
l_s(\tilde{c}_{hij}; Z, C_{-hij})+ \frac{\partial}{\partial \lambda} \log \mathcal{P}_{c_{hij} \mid C_{-hij}} (c) \bigg|_{c=\tilde{c}_{hij}}=0,
\end{equation*}
where $l_s(\tilde{c}_{hij}; Z, C_{-hij})$ is the entry of the score function, i.e. the derivative of the loglikelihood $\mathcal{L}(Z;C_1,\ldots,C_k)$, corresponding to the indices $h,i,j$.
Given prior symmetry with respect to zero, without loss of generality, we focus on $\hat{c}_{hij}>0$.
In a neighbourhood $(\hat{c}_{hij}-\varepsilon, \hat{c}_{hij}+\varepsilon)$ of the conditional maximum likelihood estimate $\hat{c}_{hij}$ of $c_{hij}$, we can approximate the score function using a Taylor expansion:
$$l_s(c; Z, C_{-hij}) = -  \mathcal{J}(\hat{c}_{hij}) \, (c - \hat{c}_{hij}) + o_\varepsilon,
$$
where $\mathcal{J}(\hat{c}_{hij})>0$ is the negative of the second derivative of $\mathcal{L}(Z;C_1,\ldots,C_k)$ with respect to $c_{hij}$ evaluated at $\hat{c}_{hij}$, and 
$o_\varepsilon$ is an approximation error term such that 
$\lim_{\varepsilon \to 0} o_\varepsilon/\varepsilon = 0$. 

For $\hat{c}_{hij}$ large enough, such that $\hat{c}_{hij}-\varepsilon>L$ with $L\gg0$, we rely on equation \eqref{eq:lim-score} for every $c$ in $(\hat{c}_{hij}-\varepsilon, \hat{c}_{hij}+\varepsilon)$, leading to the lower bound
\begin{equation*}
-\mathcal{J}(\hat{c}_{hij}) \, (c - \hat{c}_{hij}) +  f_{lb}'(c) +o_{\varepsilon}  \leq l_s(c; y)+ \frac{\partial}{\partial c} \log\{\mathcal{P}_{c_{hij} \mid C_{-hij}} (c)\}, 
\end{equation*}
where  $f_{lb}'(c)$ is a non positive continuous function with $\lim_{c \to +\infty} f_{lb}'(c) = 0$.
Let $\varepsilon$ be a function of $\hat{c}_{hij}$ such that
    $\lim_{\hat{c}_{hij} \to \infty} \varepsilon = 0$ and
    $\lim_{\hat{c}_{hij}\to \infty} f_{lb}'(\hat{c}_{hij})/ \varepsilon= 0$.   
The limit for $\hat{c}_{hij} \to \infty$ of the lower bound evaluated in $\hat{c}_{hij}-\varepsilon$ is
\begin{equation*}
 \lim_{\hat{c}_{hij} \to \infty} \mathcal{J}(\hat{c}_{hij}) \, \varepsilon +  f_{lb}'(\hat{c}_{hij}-\varepsilon) +o_{\varepsilon} = \lim_{\hat{c}_{hij} \to \infty} |\varepsilon|\, \{\mathcal{J}(\hat{c}_{hij}) +  f_{lb}'(\hat{c}_{hij}-\varepsilon)/|\varepsilon| +o_{\varepsilon}/|\varepsilon|\}.
\end{equation*}
Under the informative data assumption reported in the theorem,  $\mathcal{J}(\hat{c}_{hij})$ is of order greater or equal than $O(1)$ when $\hat{c}_{hij} \rightarrow \infty$, guaranteeing
$$
\lim_{\hat{c}_{hij} \to \infty} \mathcal{J}(\hat{c}_{hij}) +  f_{lb}'(\hat{c}_{hij}-\varepsilon)/|\varepsilon| +o_{\varepsilon}/|\varepsilon| \geq 0,
$$
such that
$
\hat{c}_{jh}-\varepsilon \leq \tilde{c}_{jh} \leq \hat{c}_{jh},
$ 
and, as a consequence,
$
\lim_{\hat{c}_{jh} \to \infty}|\tilde{c}_{jh}-\hat{c}_{jh}|=0,
$
which proves the theorem.

\end{proof}

\clearpage

\title{Supplementary Material for "Accelerated structured matrix factorization"}
\markboth{}{}
\author{}
\date{}

\setcounter{figure}{0} 
\setcounter{section}{0}
\setcounter{table}{0}
\renewcommand{\thesection}{S\arabic{section}} 
\renewcommand{\thefigure}{S\arabic{figure}} 
\renewcommand{\thetable}{S\arabic{table}} 
\renewcommand\theproposition{S\arabic{proposition}}

\maketitle

Section \ref{sec:algorithm_SM} provides a detailed description of the steps of the \XFILE algorithm presented in Section 3 of the article, and additional considerations on its implementation. Section \ref{sec:figures} contains a figure displaying the prior distribution on the number of factors $k$, for a combination of values of the parameters $\delta$ and $\alpha$.

\newpage

\section{Algorithm}\label{sec:algorithm_SM}
Let us define the following set of indices updated at every step and repetition $t$ of the loop: $\mathcal{J}^{(t)}= \{j=1,\ldots,p: v_{hj} \neq 0\}$, with $p_{\mathcal{J}}=|\mathcal{J}^{(t)}|$; $\mathcal{I}_\beta^{(t)}=\{\}$, with $n_{\mathcal{I}_\beta}=|\mathcal{I}_\beta^{(t)}|$;
$\mathcal{I}^{(t)}= \{i=1,\ldots,n: u_{ih} \neq 0\}$, with $n_{\mathcal{I}}=|\mathcal{I}^{(t)}|$M; $\mathcal{J}_\gamma^{(t)}=\{\}$, , with $n_{\mathcal{J}_\gamma}=|\mathcal{J}_\gamma^{(t)}|$.

\begin{enumerate}[label=\arabic*]
	\item  \textit{Parameter vector $\tilde{u}_h$ update.} 
	Set $\tilde{\psi}_{ih}=1$ for $i=1,\ldots,n$. Then, exploiting the
	minorize-maximize paradigm, update $\tilde{u}_h$ using the following quadratic minorant of the Student-\textit{t} loglikelihood \citep[see][for a complete presentation]{wu2010}, tangent to the current value $\tilde{u}_{h}^{(t-1)}$:
	\begin{align*}
	\sum_{i=1, j=1}^{n,p} -\left({a_\sigma+\frac{1}{2}}  \right) \bigg[
	\log\bigg\{ &1+\frac{(\tilde{z}_{ij}-\tilde{u}_{h}^{(t-1)} A_{ij})^2}{2b_\sigma}\bigg\} \\
	&+\frac{(\tilde{z}_{ij}-\tilde{u}_h A_{ij})^2-(\tilde{z}_{ij}-\tilde{u}_{h}^{(t-1)} A_{ij})^2}{2b_\sigma + (\tilde{z}_{ij}-\tilde{u}_{h}^{(t-1)} A_{ij})^2}\bigg],
	\end{align*}
	with $A_{ij} = \eta_h \frelu(x_i^\top \beta_h) \frelu(w_j^\top\gamma_h)\,v_{hj}$.
	Let $D_u^2$ denote a $np_{J}\times np_{J}$ diagonal matrix with the generic diagonal entry equal to $A^{-2}_{ij}\{2b_\sigma +(A_{ij}^{-1} \tilde{z}_{jh} -\tilde{u}^{(t-1)}_{jh})^2\}$ if and only if $j$ belongs to $J_h$ and let $\bar{z}_u$ denote a $np_J$-variate vector with generic entry $A_{ij}^{-1} \tilde{z}_{ij}$ if and only if $j$ belongs to $J_h$.
	Then, minimize 
	\begin{equation*}
	\left({a_\sigma+\frac{1}{2}}  \right)|| D_{\tilde{u}}^{-1} (\bar{z}_{\tilde{u}}- \mathbbm{1}_{np_{J}}\tilde{u}_h)||^2+\frac{||\tilde{u}_{h}||^2}{2},
	\end{equation*}
	with respect to $\tilde{u}_h$, where  $\mathbbm{1}_{np_{J}}=I_n \otimes (1,\ldots,1)^\top$ is a $np_{J}\times n$ matrix obtained as the Kronecker product between the identity matrix and a $p_{J}$-variate vector of ones.
	The optimization problem is solved by 
	\begin{equation*}
	\tilde{u}_{h}^{(t)} = \left\{\mathbbm{1}_{np_J}^\top D_{u}^{-2}\mathbbm{1}_{np_J}+\frac{1}{2(a_\sigma+0.5)}I_{n}\right\}^{-1} \mathbbm{1}_{np_J}^\top D_{u}^{-2} \bar{z}_{u}.
	\end{equation*}
	Notice that $\mathbbm{1}_{np_J}^\top D_{u}^{-2}\mathbbm{1}_{np_J}$ is a diagonal matrix with element $i$ equal to $\sum_{j \in J_h} D_{u;ij}^{-2}$, such that a low computational effort is required to perform the inversion.
	
	\item \textit{Scale $\tilde{\psi}_{h}$ update.} 
	For $i=1,\ldots,n$, set $\tilde{\psi}_{ih}^{(t)}=1$ if
	\begin{align*}	
	\hspace*{-15pt} &\log\left(\frac{\zeta_n}{1-\zeta_n}\right)> -(a_\sigma+0.5) \sum_{j=1}^{p} \left[\log\{1+\tilde{z}_{ij}^2/(2b_\sigma)\} - \log\{1+A_{ij}^2/(2b_\sigma)\} \right],\\
	&\text{with} \quad A_{ij}= \tilde{z}_{ij}-\frelu(x_i^\top \beta_h) \frelu(w_j^\top\gamma_h)\,  \tilde{u}_{ih} v_{hj} \eta_h^?
	\end{align*}
	and 0 otherwise. 
	
	\item \textit{Vector $\beta_{h}$ update.} 
	The vector $\beta_{h}$ is updated by applying a Newton-Raphson step to maximize the minorant of the Student-\textit{t} loglikelihood tangent to the current value $\beta_{h}^{(t-1)}$.
	Let $x_{\mathcal{I}_\beta}$ denote the submatrix of $x$ composed by the rows with index $i$ belonging to $\mathcal{I}_\beta^{(t)}$.
	Letting $A_{ij}=\frelu(w_j^\top\gamma_h)\,u_{ih} v_{hj} \eta_{h}$, define the $n_{\mathcal{I}_\beta} p_{\mathcal{J}}$-variate vector $\bar{z}_\beta$ with generic entry $A_{ij}^{-1} \tilde{z}_{ij}$ if and only if $i$ and $j$ belong to $\mathcal{I}^{(t)}_\beta$  $\mathcal{J}^{(t)}$, respectively, and 
	the $n_{\mathcal{I}_\beta} p_{\mathcal{J}} \times n_{\mathcal{I}_\beta} p_{\mathcal{J}}$ diagonal matrix $D_{\beta}$, where a generic entry of $D^2_{\beta}$ is $A_{ij}^{-2}[2b_\sigma + \{ A_{ij}^{-1} \tilde{z}_{ij}-\frelu(x_i^\top \beta_{h}^{(t-1)})\}]$ if and only if $i \in \mathcal{I}^{(t)}_\beta$ and $j \in \mathcal{J}^{(t)}$.
	Then, update $\beta_h^{(t)}$ setting
	\begin{equation*}
	\beta_h^{(t)}= \left\{x_{I}^\top\mathbbm{1}_{n_{\mathcal{I}_\beta} p_{\mathcal{J}}}^\top D_{\beta}^{-2}\mathbbm{1}_{n_{\mathcal{I}_\beta} p_{\mathcal{J}}}x_{I}+\frac{1}{2(a_\sigma+0.5)}I_{q_x}\right\}^{-1} \left\{ x_{I}^\top\mathbbm{1}_{n_{\mathcal{I}_\beta} p_{\mathcal{J}}}^\top D_{\beta}^{-2} \bar{z}_{\beta}+\frac{1}{2(a_\sigma+0.5)}\mu_\beta \right\},
	\end{equation*} 
	where $\mu_\beta = (1-\epsilon, 0,\ldots,0)^\top$  is the prior mean of $\beta_h$ and $\mathbbm{1}_{n_{\mathcal{I}_\beta} p_{\mathcal{J}}}= (1,\ldots,1)^\top \otimes I_{n_{\mathcal{I}_\beta}}$ is the Kronecker product of an identity matrix and a $p_{\mathcal{J}}$-variate vector of ones.
	Because of the shape of the $\frelu$ function around zero, the gradient with respect to $\beta_h$ does not exist for some points of the domain. To overcome this issue, we assume 
	\begin{equation*}
	\frac{\text{d} \,\frelu(x_i^\top \beta_h)}{\text{d}\,\beta_h} =0
	\end{equation*}
	if $\frelu(x_i^\top \beta_h)=0$, relying on the subgradient concept \citep{lange2013}.
	
	\item \textit{Vector $\tilde{v}_h$ update.} 
	Set $\tilde{\phi}_{jh}=1$ for $i=1,\ldots,n$ and  define $v_h^* = \tilde{v}_h\eta_h$, such that the prior on $v_h^*\mid \eta_h$ is the  $p$-variate Gaussian  $N_p(0, \eta_h^2)$.
	Letting $A_{ij} = \frelu(x_i^\top \beta_h) \frelu(w_j^\top\gamma_h)\,u_{ih}$, we define $\bar{z}_{v}$ as a $n_{\mathcal{I}}p$-variate vector with generic entry  $A_{ij}^{-1} \tilde{z}_{ij}$ if and only if $i$ belongs to $\mathcal{I}^{(t)}$.
	At each iteration $t$,  update $v_h^*$ with
	\begin{equation*}
	v_{h}^{*(t)} = \left\{\mathbbm{1}_{n_\mathcal{I} p}^\top D_{v^*}^{-2}\mathbbm{1}_{n_I p}+\frac{1}{2(a_\sigma+0.5)\eta_h^2}I_{n}\right\}^{-1} \mathbbm{1}_{n_I p}^\top D_{v^*}^{-2} \bar{z}_{\lambda^*},
	\end{equation*}
	where $D_{v^*}^2$ is a $n_I p\times n_I p$ diagonal matrix with a generic entry $A_{ij}^{-2}\{2b_\sigma + (A_{ij}^{-1}\tilde{z}_{ij}- v_{h}^{*(t-1)})^2\}$ if and only if $i$ belongs to $\mathcal{I}^{(t)}$.
	Finally, set $\tilde{v}_{hj}^{(t)} = v_{hj}^{*(t)}/\eta_h$.
	
	\item  \textit{Scale $\tilde{\phi}_{h}$ update.} 
	For $j=1,\ldots,p$, set $\tilde{\phi}_{jh}^{(t)}=1$ if
	\begin{align*}	
	\hspace*{-15pt} &\log\left(\frac{c_p}{1-c_p}\right)> -(a_\sigma+0.5) \sum_{j=1}^{p} \left[\log\{1+\tilde{z}_{ij}^2/(2b_\sigma)\} - \log\{1+A_{ij}^2/(2b_\sigma)\} \right],\\
	&\text{with} \quad A_{ij}= \tilde{z}_{ij}-\frelu(x_i^\top \beta_h) \frelu(w_j^\top\gamma_h)\, u_{ih} \tilde{\psi}_{ih} \tilde{v}_{hj} \eta_h
	\end{align*}
	and $0$ otherwise. 
	
	\item \textit{Vector $\gamma_{h}$ update.} 
	Let $w_{J}$ denote the submatrix of $w$ composed by the rows with index $j$ belonging to $\mathcal{J}_\gamma^{(t)}$.
	
	Letting $A_{ij}=\frelu(x_i^\top\beta_h) \,u_{ih} v_{hj} \eta_{h}$, define the $n_\mathcal{I} p_{\mathcal{J}_\gamma}$-variate vector $\bar{z}_\gamma$ with generic entry $A_{ij}^{-1} \tilde{z}_{ij}$ if and only if $i$ and $j$ belong to $\mathcal{I}^{(t)}$ and  $\mathcal{J}_\gamma^{(t)}$, respectively, and 
	the $n_\mathcal{I} p_{\mathcal{J}_\gamma} \times n_\mathcal{I} p_{\mathcal{J}_\gamma}$ diagonal matrix $D_{\gamma}$, where a generic entry of $D^2_{\gamma}$ is $A_{ij}^{-2}[2b_\sigma + \{ A_{ij}^{-1} \tilde{z}_{ij}-\frelu(w_j^\top \gamma_{h}^{(t-1)})\}]$ if and only if $i \in \mathcal{I}^{(t)}$ and $j \in \mathcal{J}^{(t)}_\gamma$.
	
	Relying on the subgradient concept, update $\gamma_h^{(t)}$ setting
	\begin{equation*}
	\gamma_h^{(t)}= \left\{w_{\mathcal{J}}^\top\mathbbm{1}_{p_{\mathcal{J}_\gamma} n_{\mathcal{I}}}^\top D_{\gamma}^{-2}\mathbbm{1}_{n_{\mathcal{J}_\gamma} n_{\mathcal{I}}}w_{\mathcal{J}}+\frac{1}{2(a_\sigma+0.5)}I_{q_w}\right\}^{-1} \left\{ w_{\mathcal{J}}^\top\mathbbm{1}_{n_{w_{\mathcal{J}_\gamma}} n_{\mathcal{I}}}^\top D_{\gamma}^{-2} \bar{z}_{\gamma}+\frac{1}{2(a_\sigma+0.5)}\mu_\gamma \right\},
	\end{equation*} 
	where $\mu_\gamma$ is the prior mean of $\gamma_h$ and $\mathbbm{1}_{n_{\mathcal{I}} p_{\mathcal{J}_\gamma}}= (1,\ldots,1)^\top \otimes I_{p_{\mathcal{J}_\gamma}}$ is the Kronecker product of an identity matrix and a $n_{\mathcal{I}}$-variate vector of ones.
	
	\item \textit{Scale $\eta_{h}$ update.} 
	The update of $\eta_h$ exploits the quantity $v_h^* = \tilde{v}_h\eta_h$ with $v_h^* \sim N_p(0, \eta_h^2)$, the full conditional distribution of $\eta_h^{-2}$ given the other parameters is
	$\text{Ga}(a_\eta+0.5p, b_\eta+0.5\sum_{j=1}^{p} v_{hj}^{*2} )$.
	Then, the value of $\eta_h^2$ maximizing the objective function is the mode of the inverse gamma distribution, i.e.,
	\begin{equation*}
	\eta_h^{2(t)} = \frac{b_\eta+0.5\sum_{j=1}^{p} v_{hj}^{*2}}{a_\eta+0.5p+1}.
	\end{equation*}	
\end{enumerate}

\begin{enumerate}[label =\arabic*]
	\setcounter{enumi}{7}
	\item \textit{Gaussian residual $\tilde{Z}$ update.} 
	For $i=1,\ldots,n$, $j=1,\ldots,p$, we set $\tilde{z}_{ij}= y_{ij}-\sum_{l=1}^{h-1} c_{lij}$ if $y_{ij}>0$. If $y_{ij}=0$, we independently update $\tilde{z}_{ij}$, setting it equal to the value that maximizes the full conditional distribution 
	\begin{equation*}
	\tilde{z}_{ij}\mid y_{ij}=0,- \sim Tt_{2a_\sigma}(\sum_{l=1}^{h} c_{lij}, b_\sigma/a_\sigma, -\infty, -\sum_{l=1}^{h-1} c_{lij}),
	\end{equation*}
	where $Tt$ indicates the truncated Student-t distribution in the interval $(-\infty,-\sum_{l=1}^{h-1} c_{lij})$.
	Then, we set $\tilde{z}_{ij}^{(t)}=\sum_{l=1}^{h} c_{lij}$ if $\sum_{l=1}^{h} c_{lij}<-\sum_{l=1}^{h-1} c_{lij}$ and $\tilde{z}_{ij}^{(t)}=-\sum_{l=1}^{h-1} c_{lij}$ otherwise.
\end{enumerate}

The algorithm structure and Steps 1,2,4,5,7 are greedy, ensuring the algorithm ascends the objective function. Then, in order to guarantee the convergence, we suggest adjusting Steps 3 and 6, relying on the Newton approximation, as follows. Given the log-posterior $l^{(t-1)}$, we perform the update as described in the algorithm if and only if the log-posterior $l^{(t-1)}$ evaluated after the step is equal or greater than $l^{(t-1)}$; otherwise, we simply move along the gradient of a small step.
We also recommend performing several random initializations of the first step of each element of $\tilde{u}_h$, $\tilde{v}_h$, $\beta_h$, and $\gamma_h$ to mitigate the risk of starting the algorithm very far from the maximum of the log-posterior, which would entail a huge number of steps to reach convergence, due to the nature of the minorize-maximize approach.

\newpage
\section{Additional figures}\label{sec:figures}

\begin{figure}[h!]
	\centering
	\includegraphics[width=7.5cm]{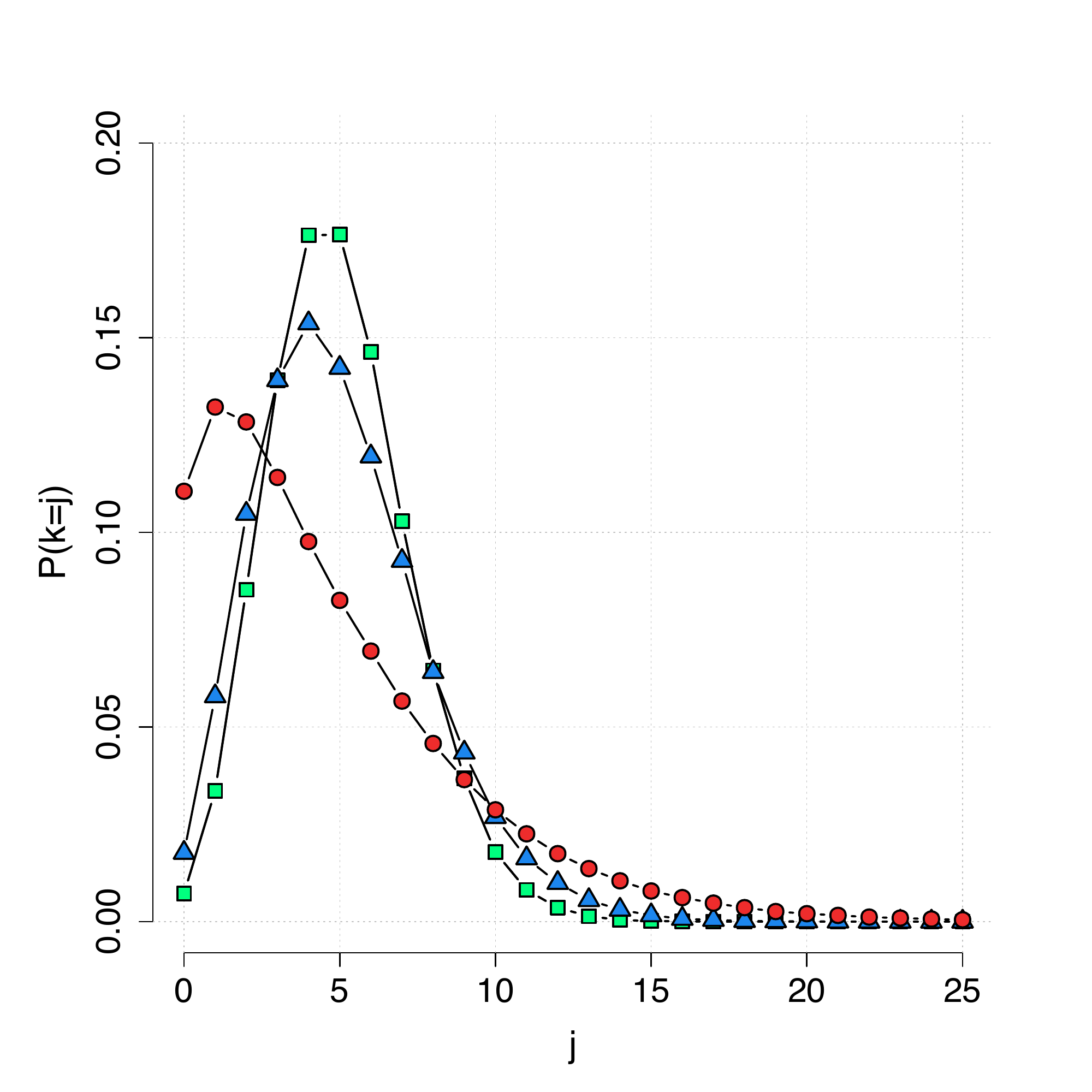} \vspace{0.5cm}\includegraphics[width=7.5cm]{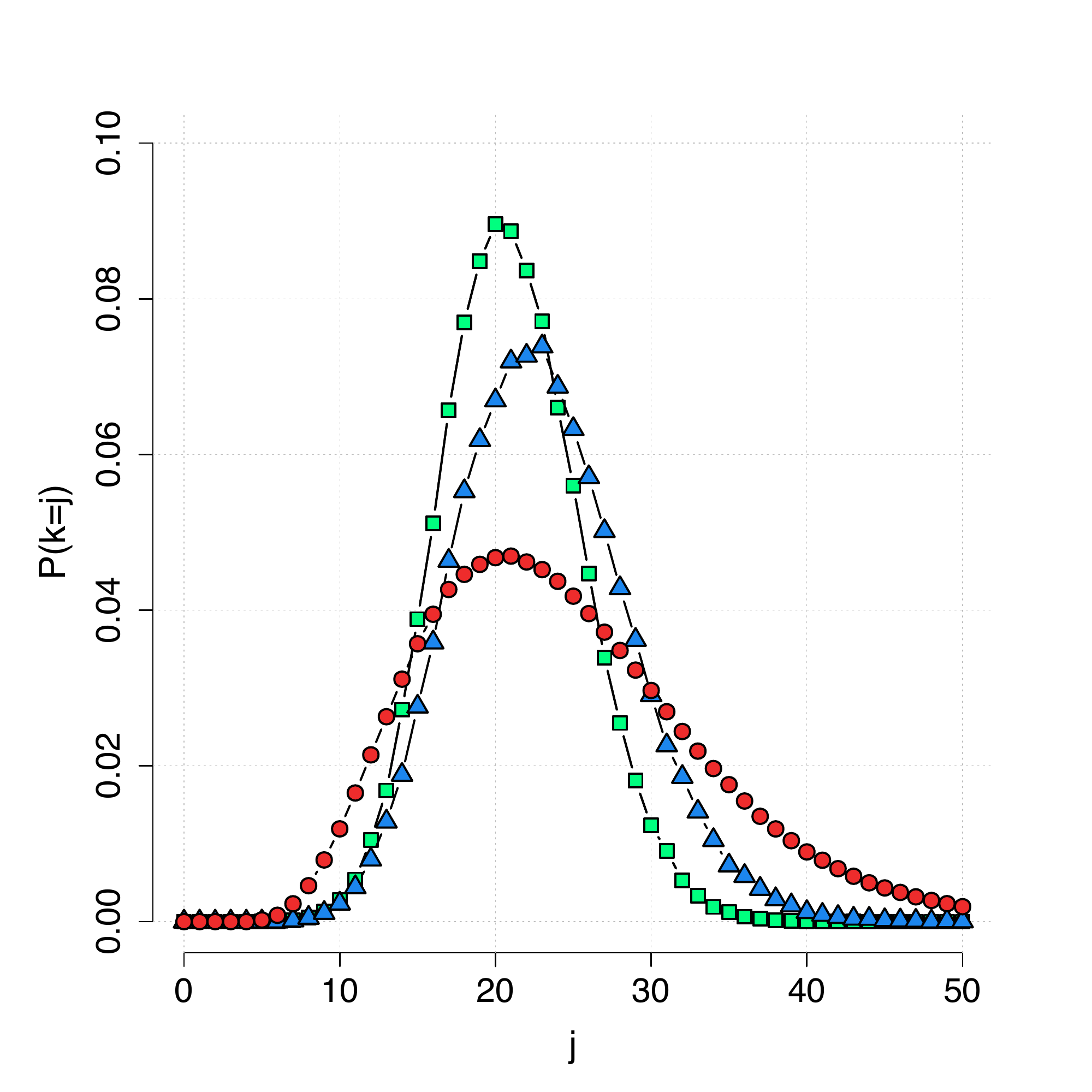}
	\caption{Left: estimated prior distribution of $k$ for three combinations of the parameters $\delta$ and $\alpha$ such that $\mathbb{E}[k]=5$, namely $(\delta,\alpha)=(0,5)$ (green squares), $(\delta,\alpha)=(0.2,2.8)$ (blue triangles) and $(\delta,\alpha)=(0.4,0.6)$ (red circles). Right: estimated prior distribution of $k$ for three combinations of the parameters $\delta$ and $\alpha$ such that $\mathbb{E}[k]=20$, namely $(\delta,\alpha)=(0,20)$ (green squares), $(\delta,\alpha)=(0.2,11.8)$ (blue triangles) and $(\delta,\alpha)=(0.4,3.6)$ (red circles).}
	\label{fig:exp_kappa}
\end{figure}


\end{document}